\documentclass{aa520}
\usepackage{txfonts}
\usepackage{graphicx}
\usepackage{natbib}
\usepackage[figuresright]{rotating}
\bibpunct{(}{)}{;}{a}{}{,}
\begin{document}

\title{The age of the Galactic thin disk from Th/Eu
nucleocosmochronology}

\subtitle{I. Determination of [Th/Eu] abundance
ratios\thanks{Based on observations collected at the European
Southern Observatory, La Silla, Chile, under the ESO programs and
the ESO-Observat\'orio Nacional, Brazil, agreement, and at the
Observat\'orio do Pico dos Dias, operated by the Laborat\'orio
Nacional de Astrof\'{\i}sica/MCT, Brazil.}}

\titlerunning{The age of the Galactic thin disk from Th/Eu
nucleocosmochronology. I}

\author{E.F. del Peloso\inst{1} \and L. da Silva\inst{1} \and
G.F. Porto de Mello\inst{2}}

\offprints{E.F. del Peloso}

\institute{Observat\'orio Nacional/MCT, Rua General Jos\'e
Cristino
77, 20921-400 Rio de Janeiro, Brazil\\
\email{epeloso@on.br, licio@on.br} \and Observat\'orio do
Valongo/UFRJ, Ladeira
do Pedro Ant\^onio 43, 20080-090 Rio de Janeiro, Brazil\\
\email{gustavo@ov.ufrj.br}}

\date{Received <date> / Accepted <date>}

\abstract{The purpose of this work is to resume investigation of
Galactic \emph{thin disk} dating using nucleocosmochronology with
Th/Eu stellar abundance ratios, a theme absent from the literature
since 1990. A stellar sample of 20~disk dwarfs/subgiants of F5 to
G8 spectral types with $-0.8\le\mathrm{[Fe/H]}\le+0.3$ was
selected. In stars with such spectral types and luminosity
classes, spectral synthesis techniques must be employed if we wish
to achieve acceptably accurate results. An homogeneous,
self-consistent set of atmospheric parameters was determined.
Effective temperatures were determined from photometric
calibrations and H$\alpha$ profile fitting; surface gravities were
obtained from $T_{\mathrm{eff}}$, stellar masses and luminosities;
microturbulence velocities and metallicities were obtained from
detailed, differential spectroscopic analysis, relative to the
Sun, using equivalent widths of \ion{Fe}{i} and \ion{Fe}{ii}
lines. Chemical abundances of the elements that contaminate the Th
and Eu spectral regions (Ti, V, Cr, Mn, Co, Ni, Ce, Nd, and Sm)
were determined through spectroscopic analysis. Abundance
uncertainties were thoroughly scrutinised, their average value --
($0.10\pm0.02$)~dex -- being found to be satisfactorily low. Eu
and Th abundances were determined by spectral synthesis of one
\ion{Eu}{ii} line (4129.72~\AA) and one \ion{Th}{ii} line
(4019.13~\AA), taking into account the detailed hyperfine
structures of contaminating Co lines, as well as the hyperfine
structure and isotope shift of the Eu line. Comparison of our
abundances with literature data shows that our results exhibit a
similar behaviour, but a considerably lower scatter (36\%~lower
for Eu, and 61\%~lower for Th). The [Th/Eu] abundance ratios thus
obtained were used, in the second paper of this series, to
determine the age of the Galactic disk.

\keywords{Galaxy: disk -- Galaxy: evolution -- Stars: late-type --
Stars: fundamental parameters -- Stars: abundances} }

\maketitle

\section{Introduction}

Modern methods of estimating the age of the Galactic thin
disk\footnote{All references to the \emph{Galactic disk} must be
regarded, in this work, as references to the \emph{thin} disk,
unless otherwise specified.}, like dating the oldest open clusters
by isochrone fitting and whites dwarfs by cooling sequences, rely
heavily on stellar evolution calculations. Nucleocosmochronology
is a dating method that makes use of only a few results of main
sequence stellar evolution models, therefore allowing a
quasi-independent verification of the afore-mentioned techniques.

Nucleocosmochronology employs the abundances of radioactive
nuclides to determine timescales for astrophysical objects and
events. The Th/Eu chronometer was first proposed by
\citet{pagel89} as a way of assessing the age of the Galactic
disk. Its main advantages are that \element[][232]Th has a
14.05~Gyr half-life, i.e., of the order of magnitude of the age
being assessed, and that Eu provides a satisfactory element for
comparison, being produced almost exclusively \citep[97\%,
according to\space][]{burrisetal00} by the same nucleosynthetic
process that produces all Th, the rapid neutron-capture process
(r-process). Its main disadvantage lies in the difficulty one
encounters when trying to determine stellar Th abundances. This
difficulty arises from the very low equivalent width (EW) of the
\ion{Th}{ii} line used, and from the fact that it is severely
blended with other much stronger lines.

The first to suggest the presence of a \ion{Th}{ii} line in the
solar spectrum, located at 4019.137~\AA, were
\citet{sitterly&king43}. \citet{severny58} was the first to
measure the EW of this line, deriving an upper limit for the solar
thorium abundance. The first confirmed Th detection and abundance
determination in a star other than the Sun was accomplished by
\citet{cowleyetal75} in HR~465, an Ap~star. Until today, only two
works available in the literature have presented Th abundances for
samples of Galactic disk stars: \citet{dasilvaetal90} and
\citeauthor{morelletal92} (1992, MKB92). Of these, only
\citeauthor{dasilvaetal90} present a chronological analysis.

This project aims at resuming investigation of Galactic disk
dating using [Th/Eu] abundance ratios, a theme absent from the
literature since 1990.\footnote{In this paper we obey the
following customary spectroscopic notations: absolute abundance
$\log\varepsilon(\mbox{A})\equiv\log_{10}(N_{\mathrm{A}}/N_{\mathrm{H}})+12.0$,
and abundance ratio
$\mbox{[A/B]}\equiv\log_{10}(N_{\mathrm{A}}/N_{\mathrm{B}})_{\mathrm{star}}
-\log_{10}(N_{\mathrm{A}}/N_{\mathrm{B}})_{\mbox{\scriptsize\sun}}$,
where $N_{\mathrm{A}}$ and $N_{\mathrm{B}}$ are the abundances of
elements A and B, respectively, in atoms~cm$^{-3}$.} In this
paper, Part~I of a series, we carried out the preliminary,
observationally-oriented steps required. First, an appropriate
stellar sample had to be elected. Selection considered the
suitability of the sample to the task of Galactic \emph{disk}
dating, with criteria like metallicity and spectral type range. In
stars with spectral types and luminosity classes adequate to our
study, which are F5--K2 dwarfs and subgiants (see
Sect.~\ref{sec:selection_criteria}), there are only one Th and one
Eu line adequate for abundance determinations. This means that
spectral synthesis techniques had to be called upon so as to
obtain acceptably accurate results. As prerequisites to the
synthesis, we determined accurate atmospheric parameters and
chemical abundances of the elements that contaminate the Th and Eu
spectral regions (Ti, V, Cr, Mn, Co, Ni, Ce, Nd, and Sm). Eu and
Th abundances were determined for all sample stars by spectral
synthesis, using our atmospheric parameters and abundances. The
[Th/Eu] abundance ratios thus obtained were used, in the second
paper of this series \citep[\space Paper~II]{delpelosoetal05b}, to
determine the age of the Galactic disk.

\section{Sample selection, observations and data reduction}

The stellar sample was defined using the Hipparcos catalogue
\citep{hipparcos}, upon which a series of selection criteria were
applied. All objects in the final sample were observed with the
Fiber-fed Extended Range Optical Spectrograph \citep[FEROS;\
][]{kauferetal99} fed by the 1.52~m European Southern Observatory
(ESO) telescope, in the ESO-Observat\'orio Nacional, Brazil,
agreement. Spectra were also obtained with a coud\'e spectrograph
fed by the 1.60~m telescope of the Observat\'orio do Pico dos Dias
(OPD), LNA/MCT, Brazil, and with the Coud\'e \'Echelle
Spectrometer (CES) fiber-fed by ESO's 3.60~m telescope and Coud\'e
Auxiliary Telescope (CAT). The selection criteria are related in
detail below, followed by the description of the observations and
their reduction.

\subsection{Selection criteria}
\label{sec:selection_criteria}

Multiple criteria were applied to the Hipparcos catalogue,
composed of 118\,218 objects. Initially, we eliminated objects
with parallaxes lower than 0.010\arcsec, in order to ensure the
minimum uncertainty in the derived bolometric magnitudes (22\,982
objects left). Only stars with declination $\delta~\le~+20\degr$
were kept, to allow the observations to be carried out in the
southern hemisphere (15\,898 objects left). Stars fainter than
visual magnitude $V~=~7.0$ were removed, so that high resolution,
high S/N ratio spectra could be obtained with relatively short
total exposures (i.e., less than 3 hours in total) on small and
medium sized telescopes (3272 objects left).

The subsequent criteria aimed at constructing the most suitable
sample for Galactic disk nucleocosmochronology. Spectral types
were restricted to the F5--K2 range because these stars have
life-times comparable to the age of the Galaxy (1744~objects
left). Only luminosity classes IV and V were allowed, to avoid
stars whose photospheric abundances had been altered by dredge-up
episodes, and to minimise non-LTE effects (925 objects left).
Since spectral analysis would be differential, we limited the
sample to stars with atmospheric parameters similar to the Sun by
selecting objects with a (B$-$V) color index in the interval
[+0.45, +0.82], so that their effective temperatures fell in the
range $T_{\mathrm{eff}\mbox{\scriptsize\sun}}\pm400$ (744 objects
left). As the ultimate objective of this work is the determination
of the age of the Galactic \emph{disk}, we eliminated stars with
[Fe/H]$<-$1.00 (252~objects left); for this purpose, we employed
average metallicities from the catalogue of
\citet{cayreldestrobeletal01}.

Stars listed as double in the Bright Star \citep{brightstar} or
Hipparcos catalogues were rejected. As a last criterium, we only
kept stars whose masses could be determined with the Geneva sets
of evolutionary tracks (\citealt{schalleretal92},
\citealt{charbonneletal93},
\citealt{schaereretal93b,schaereretal93}, and
\citealt{charbonneletal96}, hereafter collectively referred to as
Gen92/96). For this purpose, stars were required to be located
between tracks in at least two of the HR diagrams constructed for
different metallicities. This left 157 objects in the sample.

To reach a suitably sized sample, we kept only the brightest stars
to end up with 2 to 4 objects per 0.25~dex metallicity bin in the
interval $-1.00\leq\mbox{[Fe/H]}\leq+0.50$, and a total of 20
dwarfs and subgiants of F5 to G8 spectral type
(Table~\ref{tab:sample}).

\begin{table*} \caption[]{Selected stellar sample.}
\label{tab:sample}
\begin{tabular}{ l r r l c c r @{.} l r @{.} l c }
\hline \hline HD & HR & HIP & Name & R.A. & DEC &
\multicolumn{2}{c}{Parallax} & \multicolumn{2}{c}{V} &
Spectral type\\
 & & & & 2000.0 & 2000.0 & \multicolumn{2}{c}{(mas)} &\multicolumn{2}{c}{} &and\\
 & & & & (h m s) & (d m s) & \multicolumn{2}{c}{}&\multicolumn{2}{c}{}& luminosity class\\
\hline 2151 & 98 & 2021 & $\beta$ Hyi & 00 25 45 & $-$77 15 15 & 133&78 & 2&80 & G1 IV\\
9562 & 448 & 7276 &--& 01 33 43 & $-$07 01 31 & 33&71 & 5&76 & G3 V\\
16\,417 & 772 & 12\,186 & $\lambda^2$ For & 02 36 59 & $-$34 34 41
& 39&16 &
5&78 & G1 V\\
20\,766 & 1006 & 15\,330 & $\zeta^1$ Ret & 03 17 46 & $-$62 34 31
& 82&51 &
5&54 & G3--5 V\\
20\,807 & 1010 & 15\,371 & $\zeta^2$ Ret & 03 18 13 & $-$62 30 23
& 82&79 &
5&24 & G2 V\\
22\,484 & 1101 & 16\,852 & 10 Tau & 03 36 52 & +00 24 06 & 72&89 & 4&28 & F8 V\\
22\,879 &--& 17\,417 &--& 03 40 22 & $-$03 13 01 & 41&07 & 6&74 & F7--8 V\\
30\,562 & 1536 & 22\,336 &--& 04 48 36 & $-$05 40 27 & 37&73 & 5&77 & G5 V\\
43\,947 &--& 30\,067 &--& 06 19 40 & +16 00 48 & 36&32 & 6&63 & F8 V\\
52\,298 &--& 33\,495 &--& 06 57 45 & $-$52 38 55 & 27&38 & 6&94 & F5--6 V\\
59\,984 & 2883 & 36\,640 &--& 07 32 06 & $-$08 52 53 & 33&40 & 5&93 & G5--8 V\\
63\,077 & 3018 & 37\,853 & 171 Pup & 07 45 35 & $-$34 10 21 &
65&79 & 5&37 & G0
V\\
76\,932 & 3578 & 44\,075 &--& 08 58 44 & $-$16 07 58 & 46&90 &
5&86 & F7--8
IV--V\\
102\,365 & 4523 & 57\,443 &--& 11 46 31 & $-$40 30 01 & 108&23 & 4&91 & G3--5 V\\
128\,620 & 5459 & 71\,683 & $\alpha$ Cen A & 14 39 37 & $-$60 50
02 & 742&12 &
$-$0&01 & G2 V\\
131\,117 & 5542 & 72\,772 &--& 14 52 33 & $-$30 34 38 & 24&99 &
6&29 & G0--1
V\\
160\,691 & 6585 & 86\,796 & $\mu$ Ara & 17 44 09 & $-$51 50 03 &
65&46 & 5&15 &
G3 IV--V\\
196\,378 & 7875 & 101\,983 & $\phi^2$ Pav & 20 40 03 & $-$60 32 56
& 41&33 &
5&12 & F7 V\\
199\,288 &--& 103\,458 &--& 20 57 40 & $-$44 07 46 & 46&26 & 6&52 & G0 V\\
203\,608 & 8181 & 105\,858 & $\gamma$ Pav & 21 26 27 & $-$65 21 58
& 108&50 & 4&22
& F7 V\\
\hline
\end{tabular}

{References: Coordinates: SIMBAD (FK5 system); parallaxes:
Hipparcos catalogue \citep{hipparcos}; visual magnitudes: Bright
Star Catalogue \citep{brightstar} for stars with an HR~number and
SIMBAD for those without; spectral types and luminosity classes:
Michigan Catalogue of HD~Stars
\citep{michigancatalog1,michigancatalog2,michigancatalog3,michigancatalog4,michigancatalog5}
for all stars, with the exception of \object{HD~182\,572} and
\object{HD~196\,755} (Bright Star Catalogue), and
\object{HD~43\,047} \citep{fehrenbach61}.}
\end{table*}

\subsection{Checks of sample contamination}
\label{sec:kinematic}

Our sample is restricted to objects closer than 40~pc. It is a
very localised sample, considering that the Galactic thin disk has
a 290~pc scale height \citep{buseretal98}. This greatly reduces
the probability of halo stars contaminating our sample, because
the local density of halo stars is less than 0.05\%
\citep{buseretal98}. In order to reduce even further the
probability of contamination, we restricted our sample to
relatively high metallicities ($\mbox{[Fe/H]}\ge-1.00$). However,
these criteria are not perfect, because there exists an
intersection between the metallicity distributions of disk and
halo stars, i.e., there exists a non-negligible, albeit small,
number of halo stars with $\mbox{[Fe/H]}\ge-1.00$. As an
additional check of contamination, we performed a kinematic
analysis of the objects.

We developed a code to calculate the U, V, and W spatial velocity
components. This code is based on the \citet{johnson&soderblom87}
formulation and makes use of the \citet{perrymanetal98} coordinate
transformation matrix. It calculates the velocity components
relative to the Sun, and then converts them to the local standard
of rest (LSR) using the velocity of the Sun taken from
\citet{dehnen&binney98}. We adopted a right-handed coordinate
system, in which U, V, and W are positive towards the Galactic
center, the Galactic rotation and the Galactic north pole,
respectively. The code input parameters were taken from the
Hipparcos catalogue, with the exception of the radial velocities,
which were determined by us when Doppler velocity corrections were
applied to the FEROS spectra. Four of our objects
(\object{HD~22\,484}, \object{HD~22\,879}, \object{HD~76\,932},
and \object{HD~182\,572}) have radial velocities determined with
the CORAVEL spectrometers \citep{udryetal99}, which have high
precision and accuracy, and there is excellent agreement between
these and our results (linear correlation with dispersion
$\sigma=0.05$~km~s$^{-1}$). Mark that \object{HD~182\,572},
although initially selected, was not included in the final sample,
because we could not obtain Th spectra for this object;
nonetheless, we used it for comparison with \citet{udryetal99}
because we had its FEROS spectrum and its derived radial velocity
($-100.8$~km~s$^{-1}$). Distances were calculated directly from
the trigonometric parallaxes; extinction was not corrected for, as
it can be considered negligible due to the proximity of our
objects (less than 40~pc). Our results are contained in
Table~\ref{tab:kinematic_data}, along with the radial velocities
used in the calculations.

\begin{table}
\caption[]{Radial velocities (RV) and spatial velocity components
(U, V, and W) of the sample stars, in a right-handed Galactic
system and relative to the LSR. All values are in km~s$^{-1}$.}
\label{tab:kinematic_data}
\begin{tabular}{ l
r @{.} l r @{.} l r @{.} l r @{.} l } \hline \hline
HD & \multicolumn{2}{c}{RV} & \multicolumn{2}{c}{U} &  \multicolumn{2}{c}{V} & \multicolumn{2}{c}{W}\\
\hline
2151 & +23 & 7 & $-$50 & 4 $\pm$1.2 & $-$41 & 9 $\pm$0.9 & $-$24 & 5 $\pm$0.4\\
9562 & $-$14 & 0 & +1 & 4 $\pm$0.5 & $-$21 & 2 $\pm$0.8 & +20 & 2 $\pm$0.4\\
16\,417 & +11 & 8 & +32 & 1 $\pm$0.5 & $-$18 & 4 $\pm$0.7 & $-$2 & 0 $\pm$0.4\\
20\,766 & +12 & 8 & $-$60 & 5 $\pm$0.9 & $-$42 & 1 $\pm$0.9 & +22 & 8 $\pm$0.6\\
20\,807 & +12 & 1 & $-$60 & 0 $\pm$0.8 & $-$41 & 3 $\pm$0.9 & +23 & 2 $\pm$0.6\\
22\,484 & +28 & 1 & +11 & 2 $\pm$0.4 & $-$10 & 0 $\pm$0.6 & $-$34 & 8 $\pm$0.5\\
22\,879 & +120 & 7 & $-$99 & 7 $\pm$0.6 & $-$80 & 5 $\pm$1.6 & $-$37 & 7 $\pm$0.9\\
30\,562 & +77 & 3 & $-$42 & 1 $\pm$0.5 & $-$67 & 5 $\pm$1.3 & $-$13 & 9 $\pm$0.6\\
43\,947 & +41 & 6 & $-$29 & 8 $\pm$2.0 & $-$6 & 1 $\pm$0.8 & +4 & 6 $\pm$0.4\\
52\,298 & +4 & 0 & +70 & 9 $\pm$1.2 & +1 & 4 $\pm$0.6 & $-$15 & 2 $\pm$0.6\\
59\,984 & +55 & 4 & +9 & 1 $\pm$0.8 & $-$74 & 8 $\pm$1.0 & $-$21 & 8 $\pm$0.8\\
63\,077 & +106 & 2 & $-$137 & 1 $\pm$1.0 & $-$55 & 6 $\pm$0.8 & +45 & 9 $\pm$0.5\\
76\,932 & +119 & 7 & $-$38 & 1 $\pm$0.4 & $-$84 & 9 $\pm$0.7 & +76 & 7 $\pm$0.8\\
102\,365 & +17 & 4 & $-$49 & 7 $\pm$0.7 & $-$33 & 7 $\pm$0.7 & +12 & 5 $\pm$0.4\\
128\,620 & $-$23 & 9 & $-$21 & 1 $\pm$0.7 & +7 & 3 $\pm$0.9 & +19 & 8 $\pm$0.4\\
131\,117 & $-$28 & 6 & $-$50 & 2 $\pm$1.5 & $-$29 & 6 $\pm$1.8 & +17 & 4 $\pm$1.0\\
160\,691 & $-$8 & 9 & $-$3 & 6 $\pm$0.4 & $-$3 & 2 $\pm$0.6 & +3 & 2 $\pm$0.4\\
196\,378 & $-$31 & 5 & $-$55 & 2 $\pm$1.2 & $-$42 & 6 $\pm$1.2 & +5 & 9 $\pm$1.0\\
199\,288 & $-$7 & 4 & +32 & 7 $\pm$0.8 & $-$96 & 4 $\pm$1.9 & +51 & 8 $\pm$1.0\\
203\,608 & $-$29 & 4 & $-$2 & 7 $\pm$0.4 & +48 & 9 $\pm$0.6 & +12 & 6 $\pm$0.4\\
\hline
\end{tabular}
\end{table}

\begin{figure}
\resizebox{\hsize}{!}{\includegraphics*{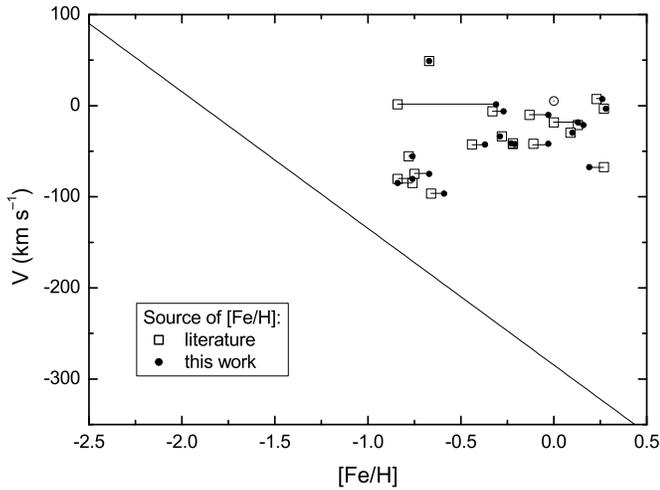}} \caption{V
vs. [Fe/H] diagram for the sample stars. The diagonal line is a
cut off between the halo and the disk populations (below and
above, respectively). Velocity components V were calculated by us.
The metallicities initially employed, represented as open squares,
were taken from the literature (Table~\ref{tab:photometry}); small
filled circles represent metallicities derived in this work.}
\label{fig:kinematics}
\end{figure}

It is not possible to discriminate the disk from the halo
population based on kinematic or metallicity criteria,
independently. Nonetheless, this discrimination can be achieved
using these criteria \emph{simultaneously}. \citet{schusteretal93}
demonstrated that halo stars whose metallicities are higher than
[Fe/H]~=~$-$1.00 have V velocity component lower than disk stars
of the same metallicity, and that it is possible to separate these
populations with a line in the V vs. [Fe/H] diagram. Analyzing
such diagram constructed for our sample stars
(Fig.~\ref{fig:kinematics}), we can see that all of them are
located far above the cut off line, indicating a highly remote
probability of one of them belonging to the halo. Note that the
metallicities used for the graph were taken from papers
individually chosen among the spectroscopic determinations that
compose the \citet{cayreldestrobeletal01} catalogue, i.e., they
were not determined by us. Even though they could have been wrong,
these determinations would have to have been \emph{overestimated}
by at least 0.52~dex for a star to cross the line, which is very
unlikely. After having determined our own metallicities
(Table~\ref{tab:adopted_atmospheric_parameters}), we confirmed
that results from the literature are overestimated by 0.08~dex at
most, when compared to our results.

According to~\citet{buseretal98}, the local density of thin disk,
thick disk, and halo stars in the Galaxy conforms to the
proportion 2000:108:1, respectively. If we take a completely
random sample of 20 stars, there is a 34.9~\% probability that it
will be totally free of thick disk stars. The probability that it
will contain 3 or more thick disk stars is low (8.0~\%).
Consequently, if there are thick disk stars in our sample, they
can be regarded as a contamination without significant bearing on
our results.

\subsection{ESO 1.52~m telescope spectra}

Observations with the ESO 1.52~m telescope were performed by the
authors in two runs, in March and August, 2001. All obtained FEROS
spectra have high nominal resolving power (R~=~48\,000),
signal-to-noise ratio ($\mbox{S/N}\ge300$ in the visible) and
coverage (3500~\AA\ to 9200~\AA\ spread over 39~\'echelle orders).
Spectra were acquired for all sample stars and for the Sun
(\object{Ganymede}), and were used for atmospheric parameter and
chemical abundance determination. Order \#9 of the FEROS spectra,
which extends from 4039~\AA\ to 4192~\AA, was used for the
determination of Eu abundances by synthesis of the \ion{Eu}{ii}
line at 4129.72~\AA.

\begin{figure}
\resizebox{\hsize}{!}{\includegraphics*{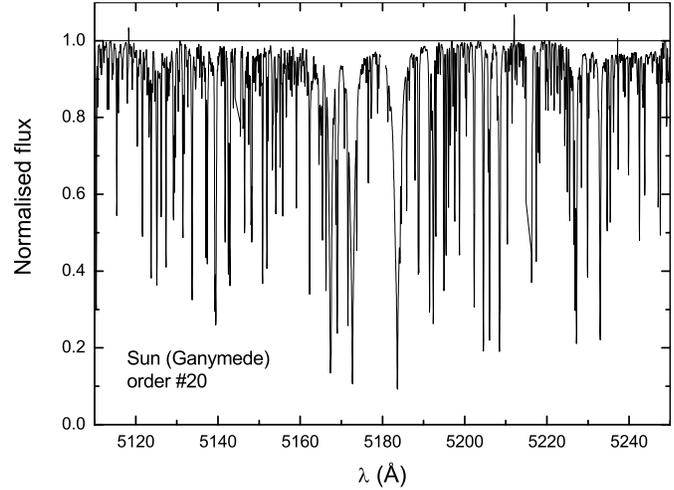}}
\caption{Example of a solar (\object{Ganymede}) normalised FEROS
spectrum (order~\#20), obtained with the ESO 1.52~m telescope.}
\label{fig:ESO_spectrum}
\end{figure}

FEROS spectra are reduced automatically by a MIDAS script during
observation, immediately after the CCD read out. Reduction is
carried out in the conventional way, applying the following steps:
bias, scattered light, and flat field corrections, extraction,
wavelength calibration (using ThAr calibration spectra), and
baricentric radial velocity correction. As a last step, the
reduction script merges all \'echelle orders to form one
continuous spectrum with a 5200~\AA\ coverage. However, the merge
process is sometimes faulty, and strong curvatures and
discontinuities are introduced in the final spectrum as result. It
becomes difficult to distinguish between the ``natural'' curvature
of the spectrum and that introduced by the merge, and the
continuum normalization procedure is rendered very insecure. We
chose to re-reduce all spectra, employing the same script used for
the automatic reduction, but eliminating the merge, so that we
could work with each order independently. Unfortunately, after the
re-reduction we found out that each individual order is plagued by
a strong discontinuity itself, which demonstrates that some
reduction steps prior to the merge are also faulty. We corrected
this by simply trimming each order and retaining only the section
beyond the discontinuity. As there are intersections in wavelength
coverage between adjacent orders, total loss of coverage caused by
the trimmings was only $\sim500$~\AA. After trimming the orders,
we corrected them for the stellar radial velocity. Stars that were
observed more than once had their spectra averaged, using their
S/N ratio as weight. Finally, the spectra were normalised by
fitting selected continuum windows with order 2 to 5 Legendre
polynomials. Figure~\ref{fig:ESO_spectrum} is an example of a
solar (\object{Ganymede}) normalised ESO spectrum.

\subsection{OPD spectra}

Observations at OPD were performed by the authors in five runs, in
May and October, 2000, and in May, August and October, 2002. The
instrumental set-up was the following:
1800~lines~$\mathrm{mm}^{-1}$ grating,
250~{\usefont{U}{psy}{m}{n}{m}}m slit, 1024
24{\usefont{U}{psy}{m}{n}{m}}m-pixel back-illuminated CCD (\#101
or \#106). All obtained spectra have moderate nominal resolving
power (R~$\sim$~22\,000) and high signal-to-noise ratio
($\mbox{S/N}\ge200$). The spectra are centered at the H$\alpha$
line (6562.8~\AA) and cover $\sim$150~\AA. Spectra were obtained
for all sample objects, with the exception of \object{HD~9562},
and for the Sun (\object{Ganymede}), and were used for
$T_{\mathrm{eff}}$ determination through profile fitting
(Sect.~\ref{sec:h_alfa_fitting}). FEROS spectra cannot be used for
this purpose because the echelle order~\#29, which contains
H$\alpha$, has a useful range of 6430~\AA\ to 6579~\AA, leaving
only 16~\AA\ at the red wing of the line. Since H$\alpha$ is very
broad, it does not reach continuum levels this close to its
center, precluding normalisation. On the hand, OPD spectra reach
75~\AA\ at each side of H$\alpha$, allowing the selection of
reliable continuum windows.

\begin{figure}
\resizebox{\hsize}{!}{\includegraphics*{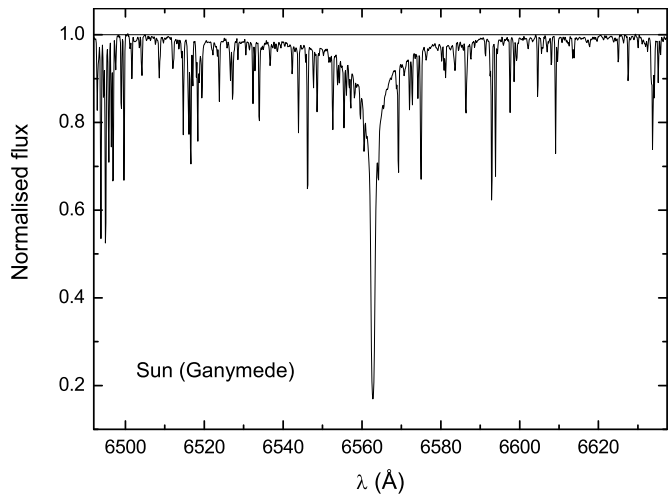}}
\caption{Example of a solar (\object{Ganymede}) normalised OPD
spectrum.} \label{fig:LNA_spectrum}
\end{figure}

Reduction was performed using the Image Reduction and Analysis
Facility (IRAF\footnote{IRAF is distributed by the National
Optical Astronomy Observatories, which are operated by the
Association of Universities for Research in Astronomy, Inc., under
cooperative agreement with the National Science Foundation.}),
following the usual steps of bias, scattered light and flat field
corrections, and extraction. Wavelength calibration was performed
individually for each star by comparing the central wavelength of
selected absorption lines with their laboratory values. This way,
Doppler corrections were not necessary, since the wavelength
calibration corrects the spectra to the rest frame. Finally,
spectra were averaged and normalised in the same way the ESO
spectra were. Figure~\ref{fig:LNA_spectrum} is an example of a
solar (\object{Ganymede}) normalised OPD spectrum.

\subsection{ESO 3.60~m telescope and CAT spectra}

Observational spectra for the Th abundance determinations were
obtained with the CES fed by the ESO 3.60~m telescope, for 16 of
the 20 sample stars and for the Sun. Observations were carried out
by the authors in January, 2002. Spectra were obtained with the
blue optical path and the high resolution image slicer (nominal
resolving power $\mbox{R}\sim235\,000$). However, unknown problems
during the observations degraded the effective resolving power
down to $\mbox{R}\sim130\,000$. The spectra were centered at the
\ion{Th}{ii} line (4019.13~\AA) and have a 27~\AA\ coverage. Due
to the low efficiency of the spectrograph at 4000~\AA, we limited
the signal-to-noise ratio to $\mbox{S/N}\sim200$ for stars with
visual magnitude $V\ge6$ and $\mbox{S/N}\sim300$ for brighter
objects.

CES spectra were also obtained with the ESO's CAT, with the aim of
determining the abundances of both Eu and Th. Observations were
carried out by the authors in August, 1998. Not all sample objects
have been observed: we obtained Eu spectra for 8 stars and the
Sun, and Th spectra for 9 stars and the Sun. All spectra have high
resolving power ($\mbox{R}=100\,000$ for Th, and
$\mbox{R}=50\,000$ for Eu), and high signal-to-noise ratio
($\mbox{S/N}\sim300$); coverage is 18~\AA.
Table~\ref{tab:available_observations} details which objects were
observed with which instruments and telescopes.

\begin{table}
\caption[]{Available observations.}
\label{tab:available_observations}
\begin{tabular}{ l @{\hspace{0em}} c @{\hspace{0.7em}} c @{\hspace{0.7em}}
c @{\hspace{0.1em}} c  c @{\hspace{0em}} c @{\hspace{0.7em}} c
@{\hspace{0.7em}} c @{\hspace{0.7em}}}
\hline \hline HD & FEROS & OPD & CES + 3.60~m & & \multicolumn{3}{c}{CES + CAT}&\\
\cline{6-8} & & & & & Europium && Thorium&\\
\hline
2151 & $\surd$ & $\surd$ & $\surd$ && $\surd$ && $\surd$&\\
9562 & $\surd$& & $\surd$ && &&&\\
16\,417 & $\surd$ & $\surd$ & $\surd$ && &&&\\
20\,766 & $\surd$ & $\surd$ & $\surd$ &&  &&$\surd$&\\
20\,807 & $\surd$ & $\surd$ & $\surd$ && $\surd$ && $\surd$&\\
22\,484 & $\surd$ & $\surd$ & $\surd$ && $\surd$ && $\surd$&\\
22\,879 & $\surd$ & $\surd$ & $\surd$ && &&&\\
30\,562 & $\surd$ & $\surd$ & $\surd$ && &&&\\
43\,947 & $\surd$ & $\surd$ & $\surd$ && &&&\\
52\,298 & $\surd$ & $\surd$ & $\surd$ && &&&\\
59\,984 & $\surd$ & $\surd$ & $\surd$ && &&&\\
63\,077 & $\surd$ & $\surd$ & $\surd$ && $\surd$ && &\\
76\,932 & $\surd$ & $\surd$ & $\surd$ && &&&\\
102\,365 & $\surd$ & $\surd$ & $\surd$ && &&&\\
128\,620 & $\surd$ & $\surd$ & $\surd$ && $\surd$ && $\surd$&\\
131\,117 & $\surd$ & $\surd$ & $\surd$ && &&&\\
160\,691 & $\surd$ & $\surd$ & && $\surd$ && $\surd$&\\
196\,378 & $\surd$ & $\surd$ & && $\surd$ && $\surd$&\\
199\,288 & $\surd$ & $\surd$ & && && $\surd$&\\
203\,608 & $\surd$ & $\surd$ & && $\surd$ && $\surd$&\\
\hline
\end{tabular}
\end{table}

\begin{figure}
\resizebox{\hsize}{!}{\includegraphics*{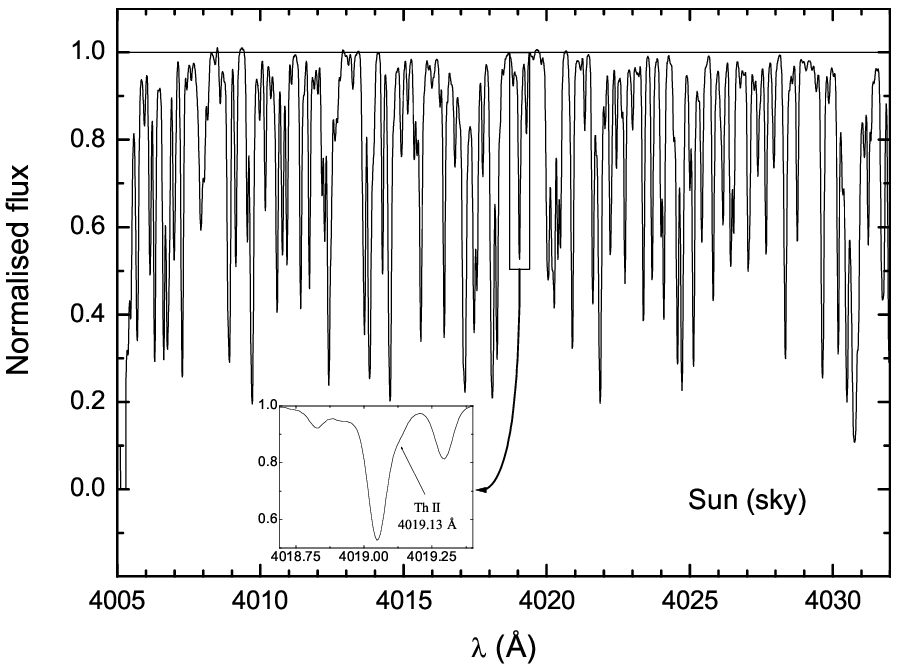}}
\caption{Example of a solar (sky) normalised CES spectrum,
observed with the ESO 3.60~m telescope, in the \ion{Th}{ii} line
region at 4019.13~\AA. The inset shows the \ion{Th}{ii} line in
greater detail.} \label{fig:ces_360_spectrum}
\end{figure}

\begin{figure}
\resizebox{\hsize}{!}{\includegraphics*{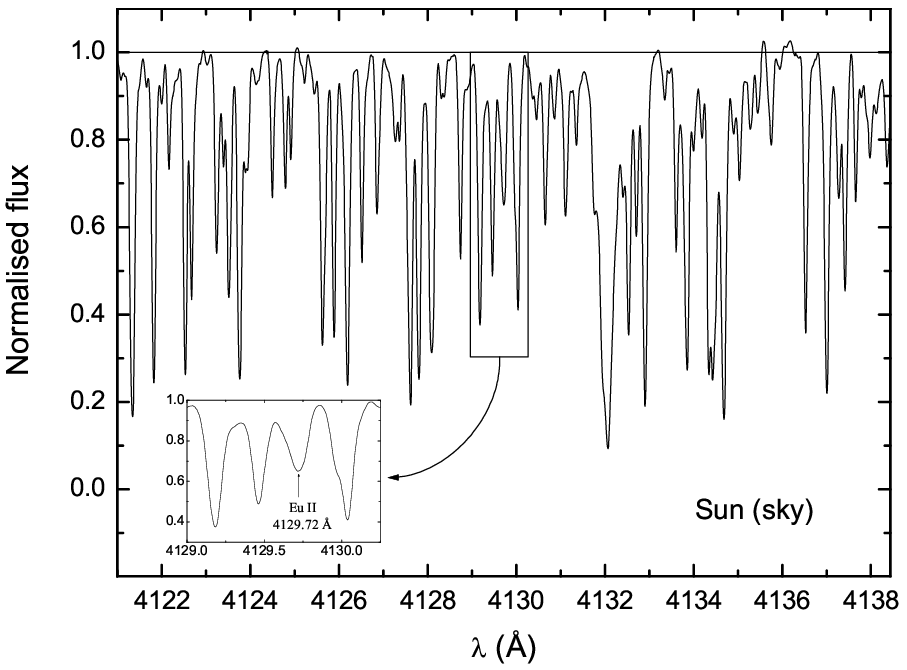}}
\caption{Example of a solar (sky) normalised CES spectrum,
observed with the ESO CAT, in the \ion{Eu}{ii} line region at
4129.72~\AA. The inset shows the \ion{Eu}{ii} line in greater
detail.} \label{fig:ces_cat_eu_spectrum}
\end{figure}

\begin{figure}
\resizebox{\hsize}{!}{\includegraphics*{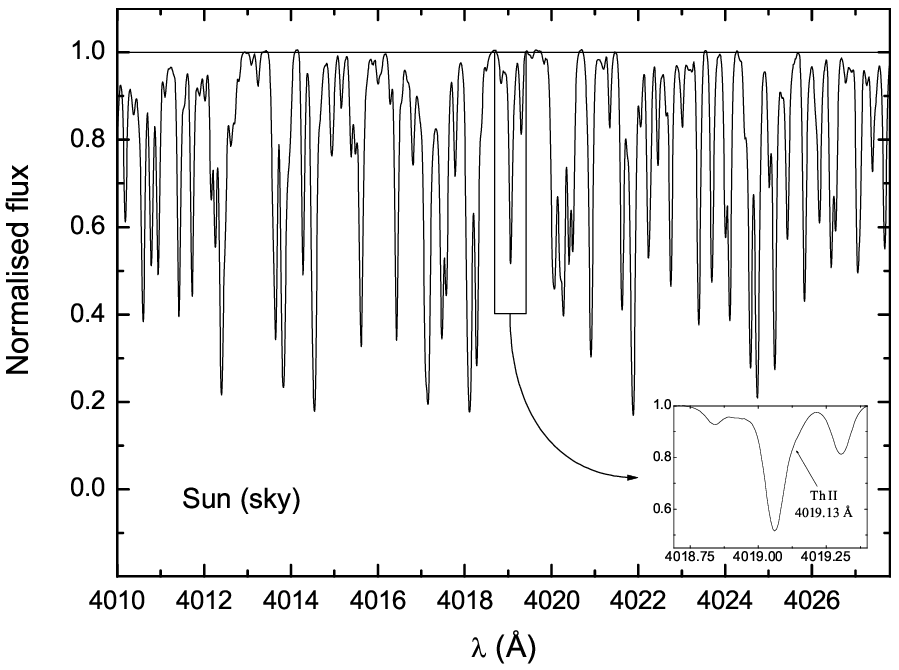}}
\caption{Example of a solar (sky) normalised CES spectrum,
observed with the ESO CAT, in the \ion{Th}{ii} line region at
4019.13~\AA. The inset shows the \ion{Th}{ii} line in greater
detail.}\label{fig:ces_cat_th_spectrum}
\end{figure}

Reduction followed the usual steps of bias and flat field
corrections, extraction and wavelength calibration (using ThAr
calibration spectra). Correction for the stellar radial velocities
was implemented by comparing the wavelengths of a number of
absorption lines with their rest frame wavelengths. Finally, the
spectra were normalised by fitting selected continuum windows with
order 2 to 5 Legendre polynomials. Stars that were observed in
more than one night did \emph{not} have spectra from different
nights averaged. Figure~\ref{fig:ces_360_spectrum} shows as an
example a solar (sky) normalised CES spectrum, observed with the
3.60~m. Examples of solar (sky) normalised CES spectra, observed
with the CAT, are presented in Fig.~\ref{fig:ces_cat_eu_spectrum}
(for the Eu region), and in Fig.~\ref{fig:ces_cat_th_spectrum}
(for the Th region).

\section{Atmospheric parameters}
\label{sec:atmospheric_parameters}

Atmospheric parameters (effective temperatures, surface gravities,
microturbulence velocities ($\xi$), and metallicities) have been
obtained through an iterative and totally self-consistent
procedure, composed of the following steps:

\begin{enumerate}
    \item Determine photometric $T_{\mathrm{eff}}$ ([Fe/H] from literature).
    \item Determine $\log g$.
    \item Determine $T_{\mathrm{eff}}$ from H$\alpha$ fitting.
    \item Set $T_{\mathrm{eff}}=(\mbox{photometric} +
    \mbox{H}\alpha)/2$.
    \item Determine {[Fe/H]} and $\xi$.
    \item Determine photometric $T_{\mathrm{eff}}$ ([Fe/H] from step~5).
    \item Set $T_{\mathrm{eff}}=(\mbox{photometric} +
    \mbox{H}\alpha)/2$.
    \item Have the parameters changed since step~8 of last iteration?\\ If
    yes, go back to step~2.
    \item END.
\end{enumerate}

Notice that a minimum of 2 iterations is required, since in step~8
we must compare the results obtained with those from the preceding
iteration. Input parameters for each step are simply the output
parameters of the preceding step. At all times, the adopted
effective temperature is the average of the two determinations
(photometric and from H$\alpha$ fitting). Owing to this, a new
average is performed each time a new determination is made
(steps~4 and 7, following $T_{\mathrm{eff}}$ determinations by
H$\alpha$ fitting and photometry, respectively). The techniques
used at each step are described in the next sections, and the
final results are presented in
Table~\ref{tab:adopted_atmospheric_parameters}.

\subsection{Model atmospheres and partition functions}
\label{sec:model_atm_part_funct}

The determination of effective temperatures through H$\alpha$
profile fitting, and of microturbulence velocities and chemical
abundances through detailed spectroscopic analysis, demands the
use of model atmospheres. Those employed by us were the plane
parallel, flux constant, LTE model atmospheres described and
discussed in detail by \citet{edvardssonetal93}. The $T(\tau)$
laws were interpolated using a code gently made available by
Monique Spite (Observatoire de Paris-Meudon, France). Throughout
this work, we have used [Fe/H] as the metallicity of the model
atmosphere. Partition functions were calculated by us using the
polynomial approximations of \citet{irwin81}.

\subsection{Effective temperature}
\label{sec:effective_temperature}

\subsubsection{Photometric calibrations}
\label{sec:photometric_calibrations}

We determined effective temperatures using the photometric
calibrations from \citet{artigo_tese_Gustavo}. These
metallicity-dependent calibrations were constructed using highly
accurate effective temperatures obtained with the infrared flux
method \citep{blackwelletal80,blackwelletal91}, and make use of
the $(B-V)$ and $(V-K)$ Johnson, $(b-y)$ and $\beta$ Str\"omgren,
and (B$_{\rm{T}}-$V$_{\rm{T}}$) Hipparcos colour indices:
\begin{displaymath}
\begin{array}{l r}
T_{\mathrm{eff}}(\mathrm{K})=7747-3016\,(\mathrm{B}-\mathrm{V})\,\{1-0.15\,\mathrm{[Fe/H]}\},
& \sigma=65\,\mathrm{K}\\
T_{\mathrm{eff}}(\mathrm{K})=8974-2880\,(\mathrm{V}-\mathrm{K})+440\,(\mathrm{V}-\mathrm{K})^2,
&\sigma=50\,\mathrm{K}\\
T_{\mathrm{eff}}(\mathrm{K})=8481-6516\,(\mathrm{b}-\mathrm{y})\,\{1-0.09\,\mathrm{[Fe/H]}\},
& \sigma=55\,\mathrm{K}\\
T_{\mathrm{eff}}(\mathrm{K}) =11\,654\,\sqrt{\beta-2.349}, & \sigma=70\,\mathrm{K}\\
T_{\mathrm{eff}}(\mathrm{K})=7551-2406\,(\mathrm{B}_{\mathrm{T}}-\mathrm{V}_{\mathrm{T}})\,\{1-0.2\,\mbox{[Fe/H]}\},
& \sigma=64\,\mathrm{K}
\end{array}
\end{displaymath}
A complete description of the derivation of these calibrations is
given in \citet{artigo_tese_Gustavo}.

Table~\ref{tab:photometry} contains photometric data from the
literature. The only sample star absent on the table is
\object{HD~128\,620} ($\alpha$~Cen~A), whose photometric data is
unreliable due to its high apparent magnitude. The metallicities
presented in the table were taken from papers individually chosen
from the \citet{cayreldestrobeletal01} catalogue, with the
exception of \object{HD~160\,691} \citep{artigo_tese_Gustavo}, and
were only used on step~1 of the atmospheric parameter
determination for each star. The $(b-y)$ colour indices taken from
\citet{gronbech&olsen76,gronbech&olsen77} and
\citet{crawfordetal70} were originally in the \citet{olsen83}
photometric system. Since the calibration used by us was based on
the \citet{olsen93} photometric system, we converted the
afore-mentioned data to this system through the relation
$(b-y)_{\mathrm{Olsen~1993}}=0.8858\,(b-y)_{\mathrm{Olsen~1983}}+0.0532$,
which is Equation~(1) from \citet{olsen93}.

Photometric temperatures have been averaged using the maximum
likelihood method, assuming that each value follows a Gaussian
probability distribution, which results in
$$\overline{T_{\rm{eff}}}=\sum_{i=1}^N
\frac{T_{\rm{eff}\,\emph{i}}}{\sigma_i^2}\left/\sum_{i=1}^N
\frac{1}{\sigma_i^2}\right.,$$ where $N$ is the number of
calibrations effectively used. We took the standard deviation of
the averages as their uncertainties. These vary from 27~K to 37~K,
and such a short range prompted us to adopt a single value
$\sigma=32~\mbox{K}$ for the average photometric
$T_{\mathrm{eff}}$ of each sample star. Take heed that these are
internal uncertainties only; for a detailed discussion, see
\citet{artigo_tese_Gustavo}. Final photometric effective
temperatures are presented in
Table~\ref{tab:adopted_atmospheric_parameters}.

\begin{table*}
\caption[]{Photometric indices and metallicities for all sample
stars. Metallicities were taken from the literature for use in the
kinematic characterization of the sample
(Sect.~\ref{sec:kinematic}) and as input for the first step in the
iterative determination of atmospheric parameters
(Sect.~\ref{sec:effective_temperature}).} \label{tab:photometry}
\begin{tabular}{l @{\hspace{2em}} c c @{\hspace{2em}} c c @{\hspace{2em}}
c c @{\hspace{2em}} c c @{\hspace{2em}} c c @{\hspace{2em}} c c}
\hline \hline HD & (B$-$V) & Ref. & (b$-$y) & Ref. & (V$-$K) &
Ref. & $\beta$ &
Ref. & (B$_{\rm{T}}-$V$_{\rm{T}}$)& Ref. & $[$Fe/H$]$ & Ref.\\
\hline
2151   & 0.62 & 1 & 0.379 & 3 & -- & --& 2.597 & 11 & -- & --& $-$0.11 & 15\\
9562   & 0.64 & 1 & 0.408 & 3 & 1.422 & 13 & 2.585 & 10 & 0.709 & 14 & +0.13 & 16\\
16\,417  & 0.66 & 1 & 0.412 & 3 & -- & --& -- & --& 0.730 & 14 & +0.00 & 17\\
20\,766  & 0.64 & 1 & 0.404 & 4 & 1.537 & 13 & 2.586 &  5 & -- & --& $-$0.22 & 18\\
20\,807  & 0.60 & 1 & 0.383 & 4 & -- & --& 2.592 &  5 & -- & --& $-$0.22 & 18\\
22\,484  & 0.58 & 1 & 0.363 & 3 & 1.363 & 13 & 2.608 &  5 & 0.626 & 14 & $-$0.13 & 19\\
22\,879  & 0.54 & 2 & 0.369 & 3 & -- & --& -- & --& 0.581 & 14 & $-$0.84 & 20\\
30\,562  & 0.62 & 1 & 0.403 & 3 & 1.410 & 13 & 2.610 &  5 & 0.709 & 14 & +0.27 & 21\\
43\,947  &-- &--& 0.377 & 6 & -- & --& 2.598 &  6 & 0.604 & 14 & $-$0.33 & 19\\
52\,298  & 0.46 & 2 & 0.320 & 3 & -- & --& -- & --& 0.500 & 14 & $-$0.84 & 22\\
59\,984  & 0.54 & 1 & 0.354 & 7 & -- & --& 2.599 &  5 & 0.566 & 14 & $-$0.75 & 23\\
63\,077  & 0.60 & 1 & 0.387 & 5 & -- & --& 2.590 &  5 & -- & --& $-$0.78 & 23\\
76\,932  & 0.53 & 1 & 0.368 & 5 & 1.410 & 13 & 2.595 &  5 & 0.556 & 14 & $-$0.76 & 16\\
102\,365 & 0.66 & 1 & 0.411 & 3 & -- & --& 2.588 &  6 & -- & --& $-$0.28 & 24\\
131\,117 & 0.60 & 1 & 0.389 & 4 & -- & --& 2.621 &  9 & 0.662 & 14 & +0.09 & 16\\
160\,691 & 0.70 & 1 & 0.433 & 3 & -- & --& -- & --& 0.786 & 14 & +0.27 & 24\\
196\,378 & 0.53 & 1 & 0.369 & 5 & -- & --& 2.609 &  5 & 0.579 & 14 & $-$0.44 & 23\\
199\,288 & 0.59 & 2 & 0.393 & 6 & -- & --& 2.588 &  6 & 0.638 & 14 & $-$0.66 & 25\\
203\,608 & 0.49 & 1 & 0.338 & 8 & 1.310 & 12 & 2.611 &  8 & 0.522 & 14 & $-$0.67 & 23\\
 \hline
\end{tabular}

References: 1~-~\citet{brightstar}; 2~-~\citet{mermilliod87};
3~-~\citet{olsen94b}; 4~-~\citet{olsen93};
5~-~\citet{gronbech&olsen76,gronbech&olsen77};
6~-~\citet{olsen83}; 7~-~\citet{manfroid&sterken87};
8~-~\citet{crawfordetal70}; 9~-~\citet{olsen&perry84};
10~-~\citet{crawfordetal66}; 11~-~\citet{hauck&mermilliod98};
12~-~\citet{koornneef83}; 13~-~\citet{dibenedetto98};
14~-~\citet{hipparcos}; 15~-~\citet{castroetal99};
16~-~\citet{grattonetal96}; 17~-~\citet{gehren81};
18~-~\citet{delpelosoetal00}; 19~-~\citet{chenetal00};
20~-~\citet{fuhrmann98}; 21~-~\citet{dasilva&portodemello00};
22~-~\citet{hartmann&gehren88}; 23~-~\citet{edvardssonetal93};
24~-~\citet{artigo_tese_Gustavo}; 25~-~\citet{axeretal94}.
\end{table*}

\subsubsection{H$\alpha$ profile fitting}
\label{sec:h_alfa_fitting}

The H$\alpha$ profiles have been studied in detail (by
\citealt{dasilva75}, \citealt{gehren81},
\citealt{fuhrmannetal93,fuhrmannetal94}, \citealt{grattonetal96},
and \citealt{cowley&castelli02}, among others), and have been
shown, in the case of cool stars, to be rather insensitive to
surface gravity, microturbulence velocity, metallicity, and mixing
length parameter variations. They are, notwithstanding, very
sensitive to the effective temperature of the atmosphere. By
comparing the observations with theoretical profiles calculated
for various effective temperatures, we can estimate this
atmospheric parameter.

Theoretical profiles were calculated with a code developed by us
from the original routines of \citet{praderie67}. This code takes
into account the convolution of radiative, Stark
\citep{vidaletal71}, Doppler and self-resonance
\citep{cayrel&traving60} broadenings.

\begin{figure}
\resizebox{\hsize}{!}{\includegraphics*{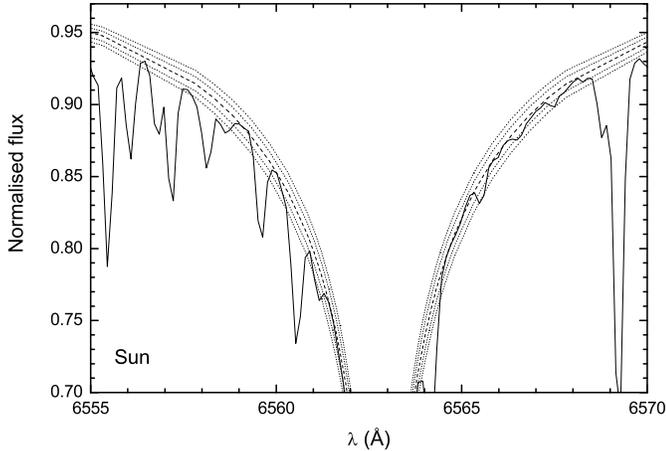}}
\caption{H$\alpha$ profile of the Sun. Dashed line is the best fit
(5767~K). Dotted lines show profiles computed when the effective
temperature was changed by $\pm$50~K and $\pm$100~K.}
\label{fig:h_alpha_fit_sun}
\end{figure}

In order to check the accuracy of the determinations, we analysed
the solar spectrum (Fig.~\ref{fig:h_alpha_fit_sun}). The adopted
solar parameters were $T_{\mathrm{eff}}=5777$~K \citep{neckel86},
$\log g=4.44$, $[\mbox{Fe/H}]=0.00$, microturbulence velocity
$\xi$ = 1.15~km~s$^{-1}$ \citep{edvardssonetal93}, and
$n(\mbox{He})/n(\mbox{H})=0.10$. The obtained temperature was
5767~K, 10~K lower than the solar value adopted by us. This small
discrepancy was corrected by adding +10~K to all stellar effective
temperatures determined by H$\alpha$ profile fitting.

Uncertainties in the determinations were estimated using the Sun
as a standard (Table~\ref{tab:h_alpha_uncertainties}). The
influence of atmospheric parameter uncertainties were evaluated
changing these parameters (one at a time) by an amount deemed
representative of their uncertainties, and then redetermining the
effective temperature. We tested the influence of the continuum
placement uncertainty, which is $\sim$0.2\%, by multiplying the
flux by 1.002, and redetermining the temperature. Since the
fitting of theoretical profile to the observation is done via a
\emph{fit-by-eye} technique, the personal judgement of who is
making the analysis introduces uncertainty to the obtained
temperature; we estimated this uncertainty to be $\sim$20~K.
Finally, some uncertainty comes from the difference between blue
and red wings, which are best fit by profiles of effective
temperatures differing by $\sim$20~K. The total uncertainty was
estimated at 43~K by the root-mean-square (RMS) of the individual
sources cited above. Final H$\alpha$ effective temperatures are
presented in Table~\ref{tab:adopted_atmospheric_parameters}.

\begin{table}
\caption[]{Uncertainty of the effective temperatures determined by
H$\alpha$ profile fitting, using the Sun as a standard. For
detailed explanation, see text.} \label{tab:h_alpha_uncertainties}
\begin{tabular}{@{} l r @{.} l c @{}}
\hline \hline Parameter & \multicolumn{2}{c}{Parameter} & T$_{\mathrm{eff}}$\\
 & \multicolumn{2}{c}{uncertainty} & uncertainty\\
 \hline
Continuum & 0 & 20~\% & 25~K\\
$\log g$ & 0 & 20~dex & 20~K\\
$[\mbox{Fe/H}]$ & 0 & 08~dex & \ \ 5~K\\
$\xi$ & 0 & 10~km~s$^{-1}$ & \ \ 5~K\\
Personal judgement & \multicolumn{2}{c}{--} & 20~K\\
Blue/red wings  & \multicolumn{2}{c}{--} & 20~K\\
\hline TOTAL & \multicolumn{2}{c}{--} & 43~K\\
 \hline
\end{tabular}
\end{table}

\subsubsection{Adopted effective temperatures}

The adopted affective temperature for each star was the arithmetic
mean of the photometric and H$\alpha$ determinations, after
convergence was achieved in the iterative procedure. The
uncertainty of the adopted $T_{\mathrm{eff}}$ was taken as the
standard deviation of the mean, resulting in 27~K.

\subsection{Surface gravity}

Surface gravities were estimated from effective temperatures,
stellar masses and luminosities, employing the known equation
\begin{equation}
\label{eqn:surface_gravity}
\log\left(\frac{g}{g_{\mbox{\scriptsize\sun}}}\right)=
\log\left(\frac{m}{m_{\mbox{\scriptsize\sun}}}\right)\
+4\log\left(\frac{T_{\mathrm{eff}}}{T_{\mathrm{eff}\mbox{\scriptsize\sun}}}\right)
+ 0.4\,(M_{\mathrm{bol}}-M_{\mathrm{bol}\mbox{\scriptsize\sun}}),
\end{equation}
in which we adopted $M_{\mathrm{bol}\mbox{\scriptsize\sun}}=4.75$
\citep{neckel86}.

\begin{figure}
\resizebox{\hsize}{!}{\includegraphics*{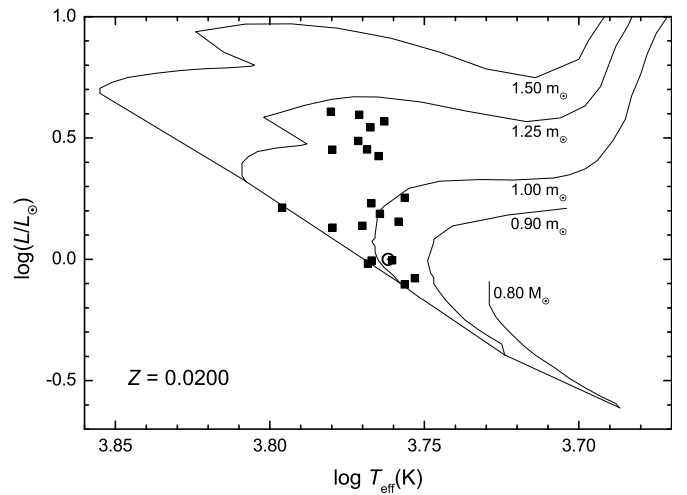}}
\caption{Example of a Gen92/96 set of evolutionary tracks, for
$Z=0.0200$. Filled squares are the sample stars, after convergence
of the atmospheric parameter determination procedure. The Sun is
represented by the symbol~$\sun$. The effective temperatures and
metallicities are the final, adopted values
(Table~\ref{tab:adopted_atmospheric_parameters}).}
\label{fig:masses}
\end{figure}

Luminosities were calculated by us from the visual magnitudes of
Table~\ref{tab:sample} and bolometric corrections from
\citet{habets&heintze81}. Stellar masses were estimated using HR
diagrams with the Gen92/96 sets of evolutionary tracks. Each
diagram was calculated by Gen92/96 for 1 of 5 different
metallicities ($Z$ = 0.0010, 0.0040, 0.0080, 0.0200, and 0.0400,
with $Z_{\mbox{\scriptsize\sun}}=0.0188)$, with tracks for 5
different masses (0.80, 0.90, 1.00, 1.25, and
1.50~$m_{\mbox{\scriptsize\sun}}$). Masses were estimated for each
star, interpolating among the tracks calculated for the
metallicities above. Then, we interpolated among metallicities to
obtain the correct mass for the star. Figure~\ref{fig:masses}
presents one such diagram, calculated for $Z=0.0200$, in which we
used the final, adopted effective temperatures and metallicities
(Table~\ref{tab:adopted_atmospheric_parameters}).

In order to check the accuracy of the determinations, we estimated
the solar mass itself in the Gen92/96 diagrams. The obtained value
was $0.97\,m_{\mbox{\scriptsize\sun}}$. This discrepancy was
corrected by adding $+0.03\,m_{\mbox{\scriptsize\sun}}$ to the
masses obtained for the sample stars.

The uncertainty of the surface gravities was obtained by error
propagation from Equation~(\ref{eqn:surface_gravity}), resulting
in the expression
$$\sigma_{\log~g}=\sqrt{\sigma_{\log(m/m_{\mbox{\scriptsize\sun}})}^2+16\,\sigma_{\log~T_{\rm{eff}}}^2};$$
analysis showed that the influence of the bolometric magnitude
uncertainty is negligible, and as such it was not taken into
consideration.

Stellar mass uncertainty is a function of the position of the star
in the HR diagram. Stars located in regions where the evolutionary
tracks are closer together will have larger errors than those
located between tracks further apart. Nevertheless, we have chosen
to adopt a single average uncertainty of
0.03~$m_{\mbox{\scriptsize\sun}}$, so as to simplify the
calculations. Taking stellar masses and effective temperatures
typical of our sample (1.00~$m_{\mbox{\scriptsize\sun}}$ and
6000~K, respectively), we have
$\sigma_{\log~T_{\rm{eff}}}=1.9\,10^{-3}$ and
$\sigma_{\log(m/m_{\mbox{\scriptsize\sun}})}=1.3\,10^{-2}$, and
consequently $\sigma_{\log~g}=0.02$~dex. This value does not take
into consideration the uncertainties intrinsic to the calculation
of evolutionary tracks, which are very difficult to assess.

\subsection{Microturbulence velocity and metallicity}

Microturbulence velocities and metallicities have been determined
through detailed, differential spectroscopic analysis, relative to
the Sun, using EWs of \ion{Fe}{i} and \ion{Fe}{ii} lines. In such
differential analysis, the Sun is treated like any other sample
star: its spectra are obtained under the same conditions, with the
same resolving power, signal-to-noise ratio, and equipment, and
are reduced by the same procedures; the same model atmospheres and
analysis codes are utilised. This way, systematic errors and
eventual non-LTE effects are reduced, because they are partially
cancelled out.

\subsubsection{Equivalent width measurements}
\label{sec:EW_measurement}

We were not interested solely in Fe lines, but also in the lines
of other elements that contaminate the spectral regions of the
\ion{Th}{ii} line in 4019.13~\AA\ (V, Cr, Mn, Co, Ni, Ce, and Nd),
and of the \ion{Eu}{ii} line in 4129.72~\AA\ (Ti, V, Cr, Co, Ce,
Nd, and Sm), whose EWs were then used for the abundance
determinations of Sect.~\ref{sec:chemical_abundances}.

An initial list of lines was constructed from
\citet{linhas_solares}, \citet{steffen85},
\citet{cayreldestrobel&bentolila89}, \citet{brown&wallerstein92},
\citet{furenlid&meylan90}, and \citet{meylanetal93}. Then, a
series of selection criteria were applied: lines with wavelengths
lower than 4000~\AA\ were removed, since the line density in this
region is very high, lowering the continuum and rendering its
normalization insecure, besides making it too difficult to find
acceptably isolated lines; lines with wavelengths greater than
7000~\AA\ were discarded, due to excessive telluric line
contamination in this region; lines with
$\mbox{EW}>110\,\mbox{m\AA}$ in the Sun, which are more sensitive
to uncertainties in microturbulence velocity due to saturation,
were eliminated; only isolated lines were kept, eliminating those
with known contaminations; lines whose EWs could not have been,
for any reason, accurately measured in the solar spectra, were
rejected.

\begin{figure}
\resizebox{\hsize}{!}{\includegraphics*{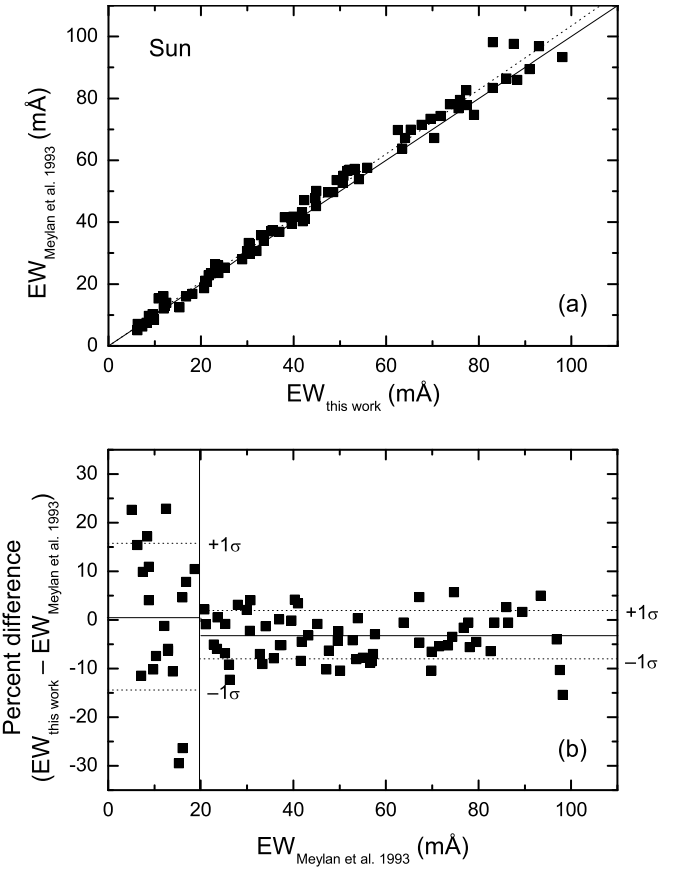}}
\caption{Panel (a): Comparison between our solar EWs and those
from \protect\citet{meylanetal93}. Solid line corresponds to a
$y=x$ relation, and the dotted one is a linear regression.
Panel~(b): Percent differences between the same EWs. Solid lines
correspond to mean differences, and the dotted ones represent the
mean differences modified by their standard deviations
($\pm1\,\sigma$).} \label{fig:EW_atlas}
\end{figure}

EWs have been measured by Gaussian profile fitting. In order to
check the measurement accuracy, we compared our solar EWs with
those from \citet{meylanetal93}. The EWs from
\citeauthor{meylanetal93} were measured by Voigt profile fitting,
using the \citet{atlas_solar} solar atlas, which has very high
resolving power and signal-to-noise ratio ($R=522\,000$ and
$S/N=3000$, between 4500~\AA\ and 6450~\AA). A linear fit between
the two data sets,
$$\mbox{EW}_{\mathrm{Meylan~et~al.~1993}}=(1.03463\pm0.00649)\,\mbox{EW}_{\mathrm{this~work}},$$
shown in panel~(a) of Fig.~\ref{fig:EW_atlas}, presents low
dispersion ($\sigma=2.9$~m\AA) and high correlation coefficient
($R=0.994$). This relation is used to correct all our EWs. The
mean percent difference between the measurements is $-2.5\%$,
presenting no dependence on the EW (Fig.~\ref{fig:EW_atlas}, panel
(b)). The larger percent differences exhibited by the weakest
lines are expected, since these are more prone to uncertainties.
More than 86\% of the lines agree at a 10\% level.

A last selection criterium was applied to the line list: given the
2.9~m\AA\ dispersion of the linear relation between our EWs and
those from \citeauthor{meylanetal93}, we rejected all Fe lines
with $\mbox{EW}<6$~m\AA\ (i.e., $\sim2\,\sigma$). This criterium
was not implemented for the lines of the other elements because
they have fewer interesting lines available, and a cut off at
6~m\AA\ would be too restrictive. The final list of lines is
composed of 65~\ion{Fe}{i}, 10~\ion{Fe}{ii}, 31~Ti, 6~V, 21~Cr,
6~Mn, 8~Co, 13~Ni, 5~Ce, 2~Nd, and 1~Sm lines, in the Sun. Not all
these lines have been used for all sample stars, since random
events, like excessive profile deformation by noise, prevented
sometimes an accurate measurement.

Table~\ref{tab:sample_ews} presents a sample of the EW data. Its
complete content, composed of the EWs of all measured lines, for
the Sun and all sample stars, is only available in electronic form
at the CDS.\footnote{Via anonymous ftp to {\tt
cdsarc.u-strasbg.fr} or via {\tt http://} {\tt
cdsweb.u-strasbg.fr/cgi-bin/qcat?}} Column~1 lists the central
wavelength (in angstroms), Column~2 gives the element symbol and
degree of ionization, Column~3 gives the excitation potential of
the lower level of the electronic transition (in eV), Column~4
presents the solar $\log gf$ derived by us, and the subsequent
Columns present the EWs, in m\AA, for the Sun and the other stars,
from \object{HD~2151} to \object{HD~203\,608} (in order of
increasing HD number).

\begin{table}
\caption[]{A sample of the EW data. The complete content of this
table is only available in electronic form at the CDS. For a
description of the columns, see text
(Sect.~\ref{sec:EW_measurement}).} \label{tab:sample_ews}
\begin{tabular}{@{} c c c c c @{\hspace{1em}} c @{\hspace{1em}}
c @{}} \hline \hline  $\lambda$~(\AA) & Element &
$\chi$~(eV) & $\log gf$ &
Sun & $\cdots$ & \object{HD~203\,608}\\
 \hline
5657.436 & \ \ion{V}{i} & 1.06 & $-$0.883 & \ \ 9.5 & $\cdots$ & \ \ 0.0\\
5668.362 & \ \ion{V}{i} & 1.08 & $-$0.920 & \ \ 8.5 & $\cdots$ & \ \ 0.0\\
5670.851 & \ \ion{V}{i} & 1.08 & $-$0.452 & 21.6 & $\cdots$ & \ \ 0.0\\
$\vdots$ & $\vdots$ & $\vdots$ & $\vdots$ & $\vdots$ & $\vdots$ & $\vdots$ \\
5427.826 & \ion{Fe}{ii} & 6.72 & $-$1.371 & \ \ 6.4 & $\cdots$ & \ \ 0.0\\
6149.249 & \ion{Fe}{ii} & 3.89 & $-$2.711 & 40.9 & $\cdots$ & 24.7\\
\hline
\end{tabular}
\end{table}

\subsubsection{Spectroscopic analysis}
\label{sec:fe_analysis}

The first step of the spectroscopic analysis consisted of
determining \emph{solar} $\log gf$ values for all measured lines.
For such determination, we employed a code written by us, based on
routines originally developed by Monique Spite. This code
calculates the $\log gf$ that an absorption line must have for its
abundance to be equal to the standard solar value from
\citet{grevesse&sauval98}, given the solar EW and lower excitation
potential of the line. For all stars, chemical abundances were
determined using these solar $\log gf$ values.

Microturbulence velocity and metallicity have been determined
through a procedure which is iterative, and which can be
summarised by the following sequence of steps:
\begin{enumerate}
    \item Determine [Fe/H] for all lines, using\\
    $[$Fe/H$]_{\mathrm{model~atmosphere}}$ = $[$Fe/H$]_{\mathrm{literature}}$.
    \item Set $[$Fe/H$]_{\mathrm{model~atmosphere}}$~=~average $[$Fe/H$]$ from lines.
    \item Determine [Fe/H] for all lines.
    \item Is $[$Fe/H$]_{\mathrm{model~atmosphere}}$~=~average $[$Fe/H$]$ from
    lines?\\ If no, go back to step 2.
    \item Draw a [Fe/H] vs. EW graph for all lines, and fit a line.
    \item Is the angular coefficient of the line negligible?\\ If
    no, modify $\xi$ and go back to step 3.
    \item END.
\end{enumerate}

In steps 1 and 3, iron abundances are calculated line-by-line
using a code very similar to the one used for the determination of
the solar $\log gf$ values. The only difference is that, in this
case, the $\log gf$ are fixed at their solar values, and an
abundance is calculated for each line. In step 6, the angular
coefficient is considered negligible when it is less than 25\% of
its own uncertainty (obtained by the fitting calculation). It is
important to note that effective temperature and surface gravity
remain constant at all times. In practice, the procedure is
executed automatically by a code, developed by \citet{dasilva00}.
This code iterates the parameters, calculating abundances and
linear fits, and making the appropriate comparisons, without human
interference.

Iron abundances obtained using \ion{Fe}{i} and \ion{Fe}{ii} lines
agree very well, their differences being always lower than
dispersion among the lines (as can be seen in
Table~\ref{tab:abundances_1}). The differences show no dependence
on any atmospheric parameter. Ionization equilibrium was not
enforced by the analysis procedure, but it was, nonetheless,
achieved.

Microturbulence velocity uncertainties have been estimated by
varying this parameter until the uncertainty of the angular
coefficient of the linear fit to the [Fe/H] vs. EW graph became
equal to the coefficient itself. We found out that the
uncertainties are metallicity-dependent, and can be divided into
three ranges:
$$ \sigma_{\xi} = \left\{  \begin{array}{l @{\mbox{,\space\space if\space}} c c c c c}
0.23~\mbox{km s}^{-1} & & & \mbox{[Fe/H]} & < & -0.7\\
0.13~\mbox{km s}^{-1} & -0.7 & \le & \mbox{[Fe/H]} & < & -0.3\\
0.05~\mbox{km s}^{-1} & -0.3 & \le & \mbox{[Fe/H]} & &
\end{array} \right.$$ This behaviour is expected, as the
dispersion between Fe lines increases for metal-poor stars.
Metallicity uncertainties were estimated in the same way that
uncertainties for the other elements were
(Sect.~\ref{sec:errors_assessment}), and are presented in
Table~\ref{tab:adopted_atmospheric_parameters}, along with the
final, adopted values of all atmospheric parameters.

\begin{table}
\caption[]{Adopted atmospheric parameters, including the
photometric and H$\alpha$ effective temperatures used to obtain
the adopted mean values, and the stellar masses used to obtain the
surface gravities. Uncertainties of photometric, H$\alpha$, and
mean effective temperatures, and of stellar masses and surface
gravities, are the same for all stars: 32~K, 43~K, 27~K,
$0.03~m_{\mbox{\scriptsize\sun}}$, and 0.02~dex, respectively.}
\label{tab:adopted_atmospheric_parameters}
\begin{tabular}{ l @{\hspace{0.47em}} c @{\hspace{0.47em}} c @{\hspace{0.3em}}
c @{\hspace{0.3em}} c @{\hspace{0.45em}} c @{\hspace{0.47em}} c
@{\hspace{0.47em}} c @{\hspace{0.47em}} c @{\hspace{0.47em}} c }
\hline \hline HD & \multicolumn{3}{c}{$T_{\mathrm{eff}}$ (K)} & &
$m_{\mbox{\scriptsize\sun}}$ & $\log g$
& [Fe/H] & $\xi$ (km~s$^{-1}$)\\
\cline{2-4} & Phot. & H$\alpha$ & MEAN & &  & & & \\
 \hline
2151 & 5909 & 5799 & 5854 & & 1.19 & 3.98 & $-$0.03 $\pm$0.09 & 1.32 $\pm$0.05 \\
9562 & 5794 & -- & 5794 && 1.23 & 3.95 & +0.16 $\pm$0.09 & 1.45 $\pm$0.05\\
16\,417 & 5821 & 5817 & 5819 && 1.15 & 4.07 & +0.13 $\pm$0.09 & 1.38 $\pm$0.05\\
20\,766 & 5696 & 5715 & 5706 && 0.95 & 4.50 & $-$0.21 $\pm$0.09 & 1.01 $\pm$0.05\\
20\,807 & 5866 & 5863 & 5865 && 0.99 & 4.48 & $-$0.23 $\pm$0.10 & 1.18 $\pm$0.05\\
22\,484 & 5983 & 6063 & 6023 && 1.16 & 4.11 & $-$0.03 $\pm$0.09 & 1.44 $\pm$0.05\\
22\,879 & 5928 & 5770 & 5849 && 0.73 & 4.34 & $-$0.76 $\pm$0.12 & 0.69 $\pm$0.23\\
30\,562 & 5883 & 5855 & 5869 && 1.18 & 4.09 & +0.19 $\pm$0.09 & 1.51 $\pm$0.05\\
43\,947 & 5940 & 5937 & 5889 && 0.95 & 4.32 & $-$0.27 $\pm$0.10 & 1.02 $\pm$0.05\\
52\,298 & 6305 & 6200 & 6253 && 1.09 & 4.41 & $-$0.31 $\pm$0.10 & 1.44 $\pm$0.13\\
59\,984 & 5968 & 5848 & 5908 && 0.97 & 3.96 & $-$0.67 $\pm$0.11 & 1.07 $\pm$0.13\\
63\,077 & 5752 & 5713 & 5733 && 0.77 & 4.15 & $-$0.76 $\pm$0.11 & 0.78 $\pm$0.23\\
76\,932 & 5874 & 5825 & 5850 && 0.82 & 4.14 & $-$0.84 $\pm$0.11 & 0.84 $\pm$0.23\\
102\,365 & 5705 & 5620 & 5663 && 0.88 & 4.43 & $-$0.29 $\pm$0.09 & 1.05 $\pm$0.05\\
128\,620 & -- & 5813 & 5813 && 1.09 & 4.30 & +0.26 $\pm$0.09 & 1.23 $\pm$0.05\\
131\,117 & 5994 & 5813 & 5904 && 1.25 & 3.96 & +0.10 $\pm$0.09 & 1.49 $\pm$0.05\\
160\,691 & 5736 & 5676 & 5706 && 1.06 & 4.19 & +0.28 $\pm$0.09 & 1.27 $\pm$0.05\\
196\,378 & 6014 & 6044 & 6029 && 1.14 & 3.97 & $-$0.37 $\pm$0.10 & 1.64 $\pm$0.13\\
199\,288 & 5785 & 5734 & 5760 && 0.82 & 4.35 & $-$0.59 $\pm$0.10 & 0.84 $\pm$0.13\\
203\,608 & 6057 & 5987 & 6022 && 0.86 & 4.31 & $-$0.67 $\pm$0.11 & 1.18 $\pm$0.23\\
 \hline
\end{tabular}
\end{table}

\section{Abundances of contaminating elements}
\label{sec:chemical_abundances}

\subsection{Abundance determination}

The abundance of the elements other than iron have been determined
using the same code as in Sect.~\ref{sec:fe_analysis}, with  the
adopted atmospheric parameters and the solar $\log gf$ values.

Adopted abundances have been obtained taking the arithmetic mean
of the abundances from all lines of each element, irrespective of
ionization state. When dispersion between lines was less than
0.10~dex, we eliminated those that deviated by more than
$2\,\sigma$ from the mean, and when it was more than 0.10~dex, we
eliminated those that deviated by more than $1\,\sigma$. The
number of lines kept is presented in Tables~\ref{tab:abundances_1}
and \ref{tab:abundances_2}, along with adopted abundances,
dispersions between the lines, and uncertainties, calculated
according to the procedure described in
Sect.~\ref{sec:errors_assessment}.

The lines of V, Mn, and Co exhibit hyperfine structures (HFSs)
strong enough that they must be taken into account, or else severe
inaccuracies would be introduced. At the resolving power achieved
by FEROS, line profiles remain Gaussian even for elements with
important HFSs, so that we can still use the EWs we measured by
Gaussian profile fitting. For V and Mn lines, we took the
wavelengths of the HFS components from the line lists of
\citet{kurucz_homepage}, whereas for the Co lines, we calculated
the wavelengths ourselves \citep{delpelosoetal05c}. These
calculations were realized using Casimir's equation
\citep{casimir_63}, and HFS laboratory data from
\citet{guthohrlein&keller90}, \citet{pickering96}, and
\citet{pickering&thorne96}. The relative intensities of the
components were obtained from the calculations of White \&
Eliason, tabulated in \citet{condon&shortley67}. A modified
version of the code, which takes HFS into consideration, was used
for the determination of the $\log gf$ values and abundances.

\begin{sidewaystable*}\begin{minipage}[t][180mm]{\textwidth}
\caption[]{Fe, Ti, V, Cr, and Mn abundances, relative to H. N is
the number of absorption lines effectively used for each abundance
determination, and $\sigma_{\mathrm{[X/H]}}$ is the standard
deviation of the lines of element~X. The uncertainties of
abundances relative to H and to Fe, calculated as described in
Sect.~\ref{sec:errors_assessment}, are also presented (as
uncert.$_{\mathrm{[X/H]}}$ and uncert.$_{\mathrm{[X/Fe]}}$,
respectively).} \label{tab:abundances_1}
\begin{tabular}{ l @{\hspace{1em}}
c @{\hspace{1em}} c @{\hspace{1em}} c @{\hspace{1em}} c
@{\hspace{1em}} c @{\hspace{1em}} c @{\hspace{1em}} c
@{\hspace{1em}} c @{\hspace{1em}} c @{\hspace{1em}} c
@{\hspace{1em}} c @{\hspace{1em}} c @{\hspace{1em}} c
@{\hspace{1em}} c @{\hspace{1em}} c } \hline \hline HD &
[\ion{Fe}{i}/H] & N & $\sigma_{\mathrm{[\ion{Fe}{i}/H]}}$ &
[\ion{Fe}{ii}/H] & N & $\sigma_{\mathrm{[\ion{Fe}{ii}/H]}}$ &
[Fe/H] & N & $\sigma_{\mathrm{[Fe/H]}}$ &
uncert.$_{\mathrm{[Fe/H]}}$ & [Ti/H] & N &
$\sigma_{\mathrm{[Ti/H]}}$
& uncert.$_{\mathrm{[Ti/H]}}$ & uncert.$_{\mathrm{[Ti/Fe]}}$\\
\hline 2151 & $-$0.03 & 54 & 0.08 & $-$0.03 & 8 & 0.05 &
-0.03 & 62 & 0.08 & 0.09 & +0.00 & 29 & 0.12 & 0.09 & 0.02\\
9562 & +0.16 & 55 & 0.06 & +0.19 & 9 & 0.04 & +0.16 & 64
& 0.06 & 0.09 & +0.15 & 29 & 0.09 & 0.09 & 0.02\\
16\,417 & +0.13 & 55 & 0.04 & +0.12 & 8 & 0.03 & +0.13 & 63 & 0.04
& 0.09 & +0.14 & 28 & 0.06 & 0.09 & 0.02\\
20\,766 & $-$0.21 & 51 & 0.05 & $-$0.21 & 8 & 0.05 & $-$0.21 & 59
& 0.05 & 0.09 & $-$0.16 & 29 & 0.06 & 0.09 & 0.02\\
20\,807 & $-$0.23 & 48 & 0.04 & $-$0.25 & 8 & 0.04 &
$-$0.23 & 56 & 0.04 & 0.10 & $-$0.18 & 28 & 0.07 & 0.09 & 0.02\\
22\,484 & $-$0.03 & 51 & 0.07 & $-$0.03 & 8 & 0.05 & $-$0.03 & 59
& 0.07 & 0.09 & +0.01 & 28 & 0.09 & 0.10 & 0.01\\
22\,879 & $-$0.75 & 38 & 0.12 & $-$0.80 & 8 & 0.08 & $-$0.76 & 46
& 0.12 & 0.10 & $-$0.46 & 27 & 0.10 & 0.09 & 0.02\\
30\,562 & +0.19 & 58 & 0.05 & +0.19 & 7 & 0.04 & +0.19 & 65 & 0.05
& 0.09 & +0.22 & 28 & 0.06 & 0.09 & 0.02\\
43\,947 & $-$0.28 & 49 & 0.06 & $-$0.22 & 8 & 0.06 & $-$0.27 & 57
& 0.07 & 0.10 & $-$0.23 & 27 & 0.10 & 0.09 & 0.02\\
52\,298 & $-$0.31 & 38 & 0.06 & $-$0.28 & 8 & 0.06 & $-$0.31 & 46
& 0.06 & 0.10 & $-$0.23 & 25 & 0.12 & 0.09 & 0.02\\
59\,984 & $-$0.67 & 43 & 0.10 & $-$0.68 & 8 & 0.11 & $-$0.67 & 51
& 0.01 & 0.11 & $-$0.48 & 28 & 0.12 & 0.08 & 0.01\\
63\,077 & $-$0.76 & 43 & 0.12 & $-$0.78 & 8 & 0.10 & $-$0.76 & 51
& 0.11 & 0.09 & $-$0.48 & 25 & 0.09 & 0.09 & 0.03\\
76\,932 & $-$0.84 & 36 & 0.11 & $-$0.84 & 8 & 0.10 & $-$0.84 & 44
& 0.11 & 0.10 & $-$0.57 & 27 & 0.08 & 0.09 & 0.02\\
102\,365 & $-$0.28 & 52 & 0.05 & $-$0.33 & 8 & 0.05 & $-$0.29 & 60
& 0.05 & 0.09 & $-$0.17 & 28 & 0.05 & 0.09 & 0.02\\
128\,620 & +0.26 & 56 & 0.06 & +0.26 & 9 & 0.04 & +0.26 & 65 &
0.06 & 0.09 & +0.29 & 27 & 0.08 & 0.09 & 0.02\\
131\,117 & +0.09 & 53 & 0.05 & +0.14 & 9 & 0.05 & +0.10 & 62 &
0.05 & 0.09 & +0.09 & 29 & 0.06 & 0.09 & 0.02\\
160\,691 & +0.27 & 52 & 0.05 & +0.33 & 8 & 0.03 & +0.28 & 60 &
0.05 & 0.09 & +0.30 & 29 & 0.07 & 0.09 & 0.02\\
196\,378 & $-$0.37 & 32 & 0.08 & $-$0.33 & 9 & 0.06 & $-$0.37 & 41
& 0.07 & 0.10 & $-$0.30 & 20 & 0.08 & 0.09 & 0.01\\
199\,288 & $-$0.59 & 37 & 0.07 & $-$0.59 & 8 & 0.06 & $-$0.59 & 45
& 0.06 & 0.10 & $-$0.38 & 22 & 0.09 & 0.09 & 0.02\\
203\,608 & $-$0.68 & 36 & 0.10 & $-$0.60 & 9 & 0.11 & $-$0.67 & 45
& 0.11 & 0.10 & $-$0.62 & 20 & 0.12 & 0.09 & 0.02\\
\hline

\end{tabular}

\vspace{0.3cm}

\begin{tabular}{ l @{\hspace{1em}} c @{\hspace{1em}} c @{\hspace{1em}} c @{\hspace{1em}} c @{\hspace{1em}}
c @{\hspace{1em}} c @{\hspace{1em}} r @{\hspace{1em}} c
@{\hspace{1em}} c @{\hspace{1em}} c @{\hspace{1em}} c
@{\hspace{1em}} c @{\hspace{1em}} c @{\hspace{1em}} c
@{\hspace{1em}} c } \hline \hline HD & [V/H] &  N &
$\sigma_{\mathrm{[V/H]}}$ & uncert.$_{\mathrm{[V/H]}}$ &
uncert.$_{\mathrm{[V/Fe]}}$ & [Cr/H] & N &
$\sigma_{\mathrm{[Cr/H]}}$ & uncert.$_{\mathrm{[Cr/H]}}$ &
uncert.$_{\mathrm{[Cr/Fe]}}$ & [Mn/H] & N &
$\sigma_{\mathrm{[Mn/H]}}$ & uncert.$_{\mathrm{[Mn/H]}}$ &
uncert.$_{\mathrm{[Mn/Fe]}}$\\ \hline 2151 & +0.02 & 4 & 0.01 &
0.10 & 0.04 & $-$0.04 & 19 & 0.06 & 0.10
& 0.02 & $-$0.09 & 6 & 0.04 & 0.08 & 0.03\\
9562 & +0.16 & 6 & 0.07 & 0.11 & 0.04 & +0.18 & 20 & 0.07 & 0.10 &
0.02 & +0.16 & 6 & 0.05 & 0.07 & 0.02\\
16\,417 & +0.18 & 6 & 0.07 & 0.11 & 0.04 & +0.14 & 19 & 0.04 &
0.10
& 0.02 & +0.14 & 6 & 0.03 & 0.07 & 0.03\\
20\,766 & $-$0.25 & 5 & 0.07 & 0.11 & 0.04 & $-$0.21 & 19 & 0.06 &
0.10 & 0.02 & $-$0.30 & 5 & 0.03 & 0.08 & 0.03\\
20\,807 & $-$0.23 & 4 & 0.03 & 0.11 & 0.04 & $-$0.23 & 19 & 0.08 &
0.11 & 0.02 & $-$0.29 & 6 & 0.03 & 0.08 & 0.03\\
22\,484 & +0.06 & 4 & 0.08 & 0.06 & 0.03 & $-$0.03 & 20 & 0.07 &
0.10 & 0.02 & $-$0.07 & 6 & 0.03 & 0.09 & 0.04\\
22\,879 & $-$0.50 & 4 & 0.15 & 0.13 & 0.05 & $-$0.69 & 13 & 0.10 &
0.11 & 0.01 & $-$0.96 & 3 & 0.10 & 0.08 & 0.03\\
30\,562 & +0.23 & 6 & 0.10 & 0.10 & 0.04 & +0.20 & 19 & 0.08 &
0.10 & 0.02 & +0.18 & 6 & 0.02 & 0.07 & 0.03\\
43\,947 & $-$0.12 & 4 & 0.08 & 0.11 & 0.04 & $-$0.27 & 20 & 0.08 &
0.11 & 0.02 & $-$0.41 & 6 & 0.09 & 0.08 & 0.03\\
52\,298 & $-$0.26 & 2 & 0.09 & 0.10 & 0.04 & $-$0.30 & 17 & 0.13 &
0.11 & 0.02 & $-$0.39 & 6 & 0.09 & 0.08 & 0.03\\
59\,984 & $-$0.44 & 4 & 0.08 & 0.13 & 0.06 & $-$0.67 & 17 & 0.13 &
0.12 & 0.01 & $-$0.83 & 6 & 0.09 & 0.09 & 0.04\\
63\,077 & $-$0.27 & 2 & 0.07 & 0.14 & 0.04 & $-$0.67 & 11 & 0.07 &
0.10 & 0.02 & $-$0.82 & 5 & 0.08 & 0.08 & 0.03\\
76\,932 & $-$0.51 & 3 & 0.17 & 0.13 & 0.05 & $-$0.76 & 10 & 0.07 &
0.11 & 0.02 & $-$1.01 & 2 & 0.06 & 0.08 & 0.03\\
102\,365 & $-$0.28 & 4 & 0.08 & 0.12 & 0.04 & $-$0.28 & 19 & 0.06
& 0.10 & 0.02 & $-$0.37 & 6 & 0.02 & 0.08 & 0.03\\
128\,620 & +0.29 & 6 & 0.10 & 0.11 & 0.04 & +0.26 & 20 & 0.06 &
0.10 & 0.02 & +0.29 & 6 & 0.04 & 0.07 & 0.02\\
131\,117 & +0.06 & 6 & 0.08 & 0.10 & 0.04 & +0.11 & 19 & 0.06 &
0.10 & 0.02 & +0.06 & 6 & 0.04 & 0.08 & 0.03\\
160\,691 & +0.25 & 4 & 0.06 & 0.12 & 0.04 & +0.28 & 19 & 0.04 &
0.10 & 0.02 & +0.32 & 6 & 0.05 & 0.05 & 0.01\\
196\,378 & -- & 0 & -- & -- & 0.04 & $-$0.38 & 14 & 0.08 & 0.11 &
0.02 & $-$0.44 & 3 & 0.01 & 0.08 & 0.03\\
199\,288 & $-$0.33 & 3 & 0.02 & 0.13 & 0.04 & $-$0.59 & 14 & 0.06
& 0.11 & 0.02 & $-$0.77 & 3 & 0.02 & 0.08 & 0.03\\
203\,608 & -- & 0 & -- & -- & 0.05 & $-$0.69 & 9 & 0.09 & 0.11 &
0.01 & $-$0.70 & 3 & 0.06 & 0.08 & 0.03\\
\hline
\end{tabular}
\vfill \end{minipage}
\end{sidewaystable*}

\begin{sidewaystable*}\begin{minipage}[t][180mm]{\textwidth}
\caption[]{The same as Table~\ref{tab:abundances_1}, but for Co,
Ni, Ce, Nd, and Sm.} \label{tab:abundances_2}
\begin{tabular} { l @{\hspace{1em}} c @{\hspace{1em}} c @{\hspace{1em}} c @{\hspace{1em}} c @{\hspace{1em}} c @{\hspace{1em}} c
@{\hspace{1em}} r @{\hspace{1em}} c @{\hspace{1em}} c
@{\hspace{1em}} c @{\hspace{1em}} c @{\hspace{1em}} c
@{\hspace{1em}} c @{\hspace{1em}} c @{\hspace{1em}} c }
\hline\hline HD & [Co/H] & N & $\sigma_{\mathrm{[Co/H]}}$ &
uncert.$_{\mathrm{[Co/H]}}$ & uncert.$_{\mathrm{[Co/Fe]}}$ &
[Ni/H] & N & $\sigma_{\mathrm{[Ni/H]}}$ &
uncert.$_{\mathrm{[Ni/H]}}$ & uncert.$_{\mathrm{[Ni/Fe]}}$ &
[Ce/H] & N & $\sigma_{\mathrm{[Ce/H]}}$ &
uncert.$_{\mathrm{[Ce/H]}}$ & uncert.$_{\mathrm{[Ce/Fe]}}$\\
\hline 2151 & +0.04 & 7 & 0.03 & 0.09 & 0.03 & $-$0.04 & 11 & 0.05
& 0.11 & 0.02 & $-$0.12 & 5 & 0.10 & 0.10 & 0.04\\
9562 & +0.18 & 7 & 0.07 & 0.09 & 0.04 & +0.20 & 11 & 0.09 & 0.10 &
0.02 & $-$0.04 & 5 & 0.05 & 0.10 & 0.04\\
16\,417 & +0.16 & 8 & 0.04 & 0.09 & 0.04 & +0.14 & 11 & 0.03 &
0.10 & 0.02 & +0.03 & 5 & 0.06 & 0.10 & 0.04\\
20\,766 & $-$0.22 & 8 & 0.06 & 0.10 & 0.04 & $-$0.22 & 11 & 0.07 &
0.11 & 0.03 & $-$0.22 & 4 & 0.06 & 0.11 & 0.04\\
20\,807 & $-$0.20 & 8 & 0.06 & 0.09 & 0.04 & $-$0.23 & 10 & 0.07 &
0.11 & 0.03 & $-$0.31 & 4 & 0.07 & 0.10 & 0.04\\
22\,484 & $-$0.01 & 8 & 0.08 & 0.05 & 0.02 & $-$0.07 & 9 & 0.05 &
0.10 & 0.01 & $-$0.14 & 5 & 0.07 & 0.07 & 0.0\\
22\,879 & $-$0.55 & 6 & 0.09 & 0.12 & 0.04 & $-$0.79 & 9 & 0.14 &
0.13 & 0.04 & $-$0.72 & 5 & 0.05 & 0.13 & 0.05\\
30\,562 & +0.24 & 8 & 0.06 & 0.09 & 0.03 & +0.23 & 11 & 0.07 &
0.10 & 0.02 & +0.14 & 5 & 0.05 & 0.10 & 0.04\\
43\,947 & $-$0.29 & 8 & 0.06 & 0.09 & 0.04 & $-$0.31 & 12 & 0.10 &
0.11 & 0.03 & $-$0.29 & 5 & 0.07 & 0.11 & 0.04\\
52\,298 & $-$0.23 & 5 & 0.08 & 0.09 & 0.03 & $-$0.38 & 5 & 0.05 &
0.11 & 0.03 & $-$0.27 & 4 & 0.09 & 0.10 & 0.04\\
59\,984 & $-$0.54 & 6 & 0.10 & 0.13 & 0.05 & $-$0.71 & 6 & 0.10 &
0.14 & 0.04 &
$-$0.81 & 5 & 0.04 & 0.14 & 0.05\\
63\,077 & $-$0.53 & 6 & 0.07 & 0.12 & 0.04 & $-$0.62 & 7 & 0.09 &
0.12 & 0.05 & $-$0.80 & 5 & 0.03 & 0.13 & 0.05\\
76\,932 & $-$0.61 & 4 & 0.02 & 0.11 & 0.04 & $-$0.80 & 7 & 0.09 &
0.12 & 0.04 & $-$0.81 & 5 & 0.08 & 0.13 & 0.05\\
102\,365 & $-$0.23 & 8 & 0.05 & 0.10 & 0.04 & $-$0.29 & 10 & 0.05
& 0.11 & 0.03 & $-$0.38 & 4 & 0.08 & 0.11 & 0.04\\
128\,620 & +0.30 & 8 & 0.06 & 0.09 & 0.04 & +0.28 & 9 & 0.04 &
0.10 & 0.02 & +0.15 & 5 & 0.08 & 0.10 & 0.04\\
131\,117 & +0.05 & 8 & 0.04 & 0.08 & 0.03 & +0.12 & 12 & 0.04 &
0.11 & 0.02 & $-$0.03 & 5 & 0.05 & 0.09 & 0.04\\
160\,691 & +0.29 & 8 & 0.04 & 0.09 & 0.04 & +0.34 & 11 & 0.04 &
0.09 & 0.02 & +0.18 & 5 & 0.06 & 0.09 & 0.04\\
196\,378 & $-$0.31 & 2 & 0.17 & 0.09 & 0.04 & $-$0.40 & 8 & 0.11 &
0.12 & 0.03 & $-$0.41 & 3 & 0.03 & 0.11 & 0.04\\
199\,288 & $-$0.39 & 4 & 0.05 & 0.11 & 0.04 & $-$0.63 & 6 & 0.05 &
0.12 & 0.04 & $-$0.66 & 3 & 0.07 & 0.12 & 0.05\\
203\,608 & $-$0.18 & 3 & 0.09 & 0.11 & 0.04 & $-$0.74 & 5 & 0.09 &
0.12 & 0.03 & $-$0.73 & 4 & 0.09 & 0.12 & 0.05\\
\hline
\end{tabular}

\vspace{0.3cm}

\begin{tabular}{ l @{\hspace{1em}}
c @{\hspace{1em}} c @{\hspace{1em}} c @{\hspace{1em}} c
@{\hspace{1em}} c @{\hspace{1em}} c @{\hspace{1em}} c
@{\hspace{1em}} c } \hline \hline HD & [Nd/H] & N &
uncert.$_{\mathrm{[Nd/H]}}$ & uncert.$_{\mathrm{[Nd/Fe]}}$ &
[Sm/H] & N &
uncert.$_{\mathrm{[Sm/H]}}$ & uncert.$_{\mathrm{[Sm/Fe]}}$\\
\hline 2151 & $-$0.03 & 2 & 0.11 & 0.04 & $-$0.24 & 1 & 0.11
& 0.04\\
9562 & +0.06 & 2 & 0.12 & 0.04 & +0.06 & 1 & 0.11 & 0.04\\
16\,417 & +0.06 & 2 & 0.12 & 0.04 & +0.06 & 1 & 0.11 & 0.04\\
20\,766 & $-$0.10 & 2 & 0.12 & 0.05 & $-$0.11 & 1 & 0.12 & 0.04\\
20\,807 & $-$0.12 & 2 & 0.12 & 0.05 & $-$0.19 & 1 & 0.11 & 0.04\\
22\,484 & +0.04 & 2 & 0.08 & 0.04 & $-$0.13 & 1 & 0.07 & 0.04\\
22\,879 & $-$0.29 & 2 & 0.14 & 0.06 & $-$0.49 & 1 & 0.13 & 0.05\\
30\,562 & +0.16 & 2 & 0.11 & 0.04 & +0.14 & 1 & 0.11 & 0.04\\
43\,947 & $-$0.17 & 2 & 0.12 & 0.05 & $-$0.28 & 1 & 0.11 & 0.04\\
52\,298 & $-$0.10 & 2 & 0.12 & 0.05 & $-$0.29 & 1 & 0.11 & 0.04\\
59\,984 & $-$0.47 & 2 & 0.15 & 0.06 & $-$0.74 & 1 & 0.14 & 0.05\\
63\,077 & $-$0.40 & 2 & 0.14 & 0.06 & $-$0.85 & 1 & 0.13 & 0.05\\
76\,932 & $-$0.44 & 2 & 0.14 & 0.06 & $-$0.93 & 1 & 0.13 & 0.05\\
102\,365 & $-$0.13 & 2 & 0.13 & 0.05 & $-$0.23 & 1 & 0.12 & 0.05\\
128\,620 & +0.28 & 2 & 0.12 & 0.04 & +0.26 & 1 & 0.11 & 0.04\\
131\,117 & +0.10 & 2 & 0.11 & 0.04 & $-$0.11 & 1 & 0.10 & 0.04\\
160\,691 & +0.14 & 2 & 0.12 & 0.03 & +0.28 & 1 & 0.12 & 0.04\\
196\,378 & -- & 0 & -- & -- & $-$0.24 & 1 & 0.11 & 0.05\\
199\,288 & $-$0.32 & 1 & 0.13 & 0.05 & $-$0.39 & 1 & 0.13 & 0.05\\
203\,608 & -- & 0 & -- & -- & $-$0.62 & 1 & 0.12 & 0.05\\
\hline
\end{tabular}
\vfill\end{minipage}
\end{sidewaystable*}

\subsection{Uncertainty assessment}
\label{sec:errors_assessment}

Chemical abundances determined using EWs are subject to three
sources of uncertainty: atmospheric parameters, EWs, and model
atmospheres. If these sources were independent from one another,
the total uncertainty could be estimated  by their RMS. However,
there is no such independency, as can be clearly perceived when
the iterative character of the analysis is called to mind. In this
circumstance, RMS yields an estimate of the \emph{maximum}
uncertainty. Given the extreme complexity of the interdependency
between the sources of uncertainty, we did not attempt to
disentangle them, and used the RMS instead.

\begin{figure}
\resizebox{\hsize}{!}{\includegraphics*{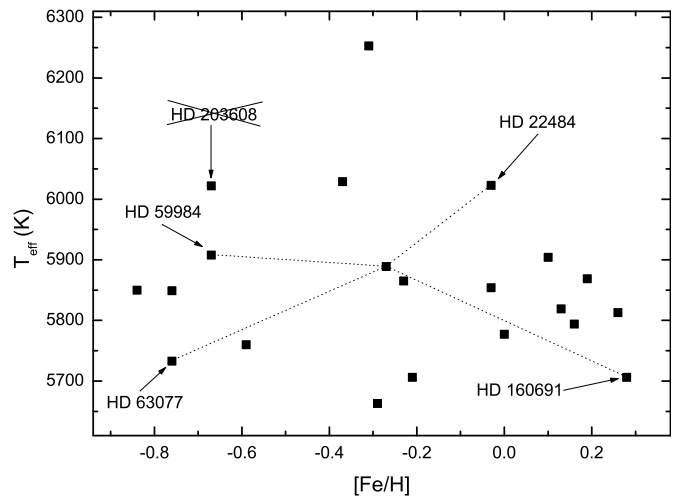}}
\caption{Stars used as uncertainty standards are labelled.
\object{HD~203\,608} was not chosen when assessing the
uncertainties of contaminating elements, because it lacks V and Nd
abundance determinations; however, it was chosen as the hot,
metal-poor standard when assessing Eu and Th uncertainties. The
effective temperatures and metallicities are the adopted ones.
Dotted lines are an example of the weights used for uncertainty
determination in stars not chosen as standards -- see text for
details.} \label{fig:error_stars}
\end{figure}

We chose four stars as standards for the assessment, with extreme
effective temperatures and metallicities: \object{HD~160\,691}
(cool and metal-rich), \object{HD~22\,484} (hot and metal-rich),
\object{HD~63\,077} (cool and metal-poor), and \object{HD~59\,984}
(hot and metal-poor). \object{HD~203\,608} is as metal-poor as,
but hotter than \object{HD~59\,984}, and would apparently be
better suited as a standard star; yet we did not use it, because
it lacks measured V and Nd lines, and thus could not be used to
estimate the uncertainties for these elements
(Fig.~\ref{fig:error_stars}).

Initially, we determined the iron abundance uncertainty. For this
purpose, we evaluated the influence of all uncertainty sources,
independently. The influence of effective temperature, surface
gravity and microturbulence velocity was obtained by recalculating
the Fe abundances using the adopted atmospheric parameters
modified, one at a time, by their own uncertainties: +27~K,
+0.02~dex, and +0.05~km~s$^{-1}$ (metal-rich stars) and
+0.23~km~s$^{-1}$ (metal-poor stars), respectively. The difference
between the adopted Fe abundance and the one obtained with the
modified parameter is our estimate of the influence of the
parameter on the total uncertainty.

\begin{figure}
\resizebox{\hsize}{!}{\includegraphics*{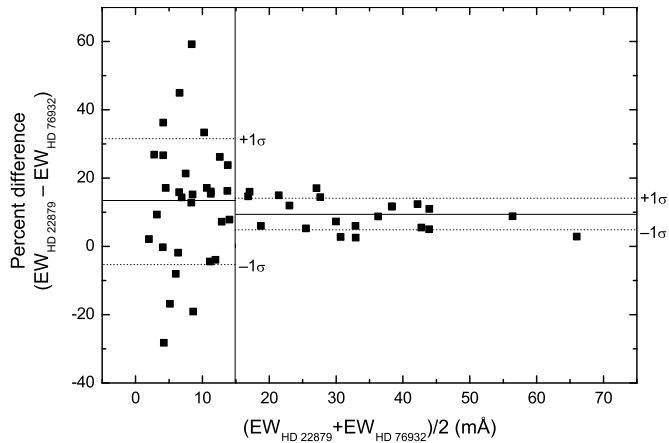}}
\caption{Percent differences between the EWs of
\object{HD~22\,879} and \object{HD~76\,932}. Solid lines
correspond to mean differences, and the dotted ones represent the
mean differences modified by their standard deviations
($\pm1\,\sigma$).} \label{fig:EW_poor_stars}
\end{figure}

The influence of EWs was determined by recalculating the
abundances using EWs modified by percentages obtained when
comparing our measurements with those from \citet{meylanetal93}
(Fig.~\ref{fig:EW_atlas}). This comparison allows us to estimate
the influence of the EW measurement per se, as well as the
influence of continuum placement. The percent difference between
the two sets of measurements presents a 15\% dispersion for lines
with $\mbox{EW}<20$~m\AA, and 5\% for those with
$\mbox{EW}>20$~m\AA. The Fe EWs were, then, modified by these
amounts. However, this was only done for the two metal-rich
standard stars. For the metal-poor stars, the influence of the
continuum placement is stronger than that estimated by comparing
solar EWs, because EWs are considerably lower. On the other hand,
it is easier to find good continuum windows, because they are
wider and more numerous. So, we compared the EWs of
\object{HD~22\,879} and \object{HD~76\,932}, which are metal-poor
stars with very similar atmospheric parameters (specially their
effective temperatures, which are virtually identical). The high
similarity of atmospheric parameters mean that differences in EWs
are caused almost exclusively by the uncertainties in the EW
measurements and continuum placement. Only \ion{Fe}{i} lines were
compared, in order to minimize the effect of the stars having
slightly different surface gravities. The EW percent difference
between the two stars present 19\% dispersion for lines with
$\mbox{EW}<15$~m\AA, and 5\% for those with $\mbox{EW}>15$~m\AA\
(Fig.~\ref{fig:EW_poor_stars}); the Fe EWs of the metal-poor
standard stars were modified by these values.

$\mbox{Log }gf$ are also a source of uncertainty. But, since they
are derived from solar EWs, their contribution is the same as that
from the EWs themselves. That is to say, in practice we take into
account the influence of EWs twice, so as to account for the solar
EWs per se, and for the $\log gf$ values.

The influence of metallicity in the Fe abundance uncertainty is
the uncertainty itself. This means that the Fe abundance
uncertainty determination is, rigorously speaking, an iterative
process. As a first step, Fe abundances were modified by the
average dispersion \emph{between lines} of all stars (0.07~dex).
As can be seen in Table~\ref{tab:errors}, the influence of
metallicity is very weak. The contribution of this source would
have to be 0.04~dex higher for the total uncertainty to be raised
just 0.01~dex. As a second step, we should have modified the Fe
abundances by the total Fe uncertainty just obtained, on the first
step, by the RMS of all sources (including the metallicity
itself). But, given the low contribution of this source, we
stopped at the first step.

The total Fe uncertainties for the standard stars were obtained by
the RMS of the six contributions: effective temperatures, surface
gravities, microturbulence velocities, EWs, $\log gf$, and
metallicities. Total uncertainties for the other stars were
obtained by weighted average of the standard stars values, using
as weight the reciprocal of the distance of the star to each
standard star, in the $T_{\mathrm{eff}}$ vs. [Fe/H] plane. In
Fig.~\ref{fig:error_stars} we see an example, in which the star at
the center uses the reciprocal of the lengths of the dotted lines
as weights.

Uncertainties for the abundances of the other elements were
determined the same way as for Fe. Although we determined
individual [Fe/H] uncertainties for each star, we used a sole
value of 0.10~dex when estimating the metallicity contribution to
the uncertainties of the other elements. This can be justified by
the low dispersion between [Fe/H] uncertainties, and by the very
low sensitivity of the abundances to the metallicity of the model
atmosphere. Table~\ref{tab:errors} presents contributions from the
individual sources, total uncertainties, and some averages for the
standard stars, relative to H. Values relative to Fe can be
obtained simply by subtracting the [Fe/H] value from the
[element/H] value (e.g.,
$\mbox{uncert.}_{\mathrm{[V/Fe]}}=\mbox{uncert.}_{\mathrm{[V/H]}}
-\mbox{uncert.}_{\mathrm{[Fe/H]}}$).

\begin{sidewaystable*}\begin{minipage}[t][180mm]{\textwidth}
\caption[]{[element/H] abundance uncertainties for the standard
stars. Last column contains the average uncertainty for each row,
and the last row contains\\ the average total uncertainty for the
four standard stars.} \label{tab:errors}
\begin{tabular} { c r @{\space} r c r @{.} l  r @{.} l r @{.} l r @{.} l r @{.} l
r @{.} l r @{.} l r @{.} l r @{.} l r @{.} l r @{\space} l }
\hline\hline Parameter & \multicolumn{2}{c}{$\Delta$Parameter} &
HD & \multicolumn{2}{c}{[Fe/H]} & \multicolumn{2}{c}{[Ti/H]} &
\multicolumn{2}{c}{[V/H]} & \multicolumn{2}{c}{[Cr/H]} &
\multicolumn{2}{c}{[Mn/H]} & \multicolumn{2}{c}{[Co/H]} &
\multicolumn{2}{c}{[Ni/H]} & \multicolumn{2}{c}{[Ce/H]} &
\multicolumn{2}{c}{[Nd/H]} & \multicolumn{2}{c}{[Sm/H]} &
\multicolumn{2}{c}{Average}\\
\hline & & & 160\,691 & +0 & 01 & +0 & 02 & +0 & 03 & +0 & 02 & +0
& 02 & +0 & 02 & +0 & 02 &
+0 & 00 & +0 & 01 & +0 & 01 & +0.02 & $\pm$0.01\\
& & & 22\,484 & +0 & 02 & +0 & 02 & +0 & 03 & +0 & 02 & +0 & 03 &
+0 & 03 & +0 & 02 &
+0 & 01 & +0 & 01 & +0 & 01 & +0.02 & $\pm$0.01\\
\raisebox{1.5ex}[0pt]{$T_{\mathrm{eff}}$} &
\multicolumn{2}{c}{\raisebox{1.5ex}[0pt]{+27~K}} & 63\,077 & +0 &
02 & +0 & 02 & +0 & 03 & +0 & 01 & +0 & 02 & +0 & 02 & +0 & 02 &
+0 & 01 & +0 & 01 & +0 & 01 & +0.02 & $\pm$0.01\\
& & & 59\,984 & +0 & 02 & +0 & 02 & +0 & 03 & +0 & 02 & +0 & 03 &
+0 & 02 & +0 & 02 &
+0 & 01 & +0 & 02 & +0 & 01 & +0.02 & $\pm$0.01\\
\hline & & & 160\,691 & +0 & 00 & +0 & 00 & +0 & 00 & +0 & 00 & +0
& 00 & +0 & 00 & +0 & 00 &
+0 & 01 & +0 & 01 & +0 & 01 & +0.00 & $\pm$0.00\\
& & & 22\,484 & +0 & 00 & +0 & 00 & +0 & 01 & +0 & 00 & +0 & 00 &
+0 & 01 & +0 & 00 &
+0 & 01 & +0 & 01 & +0 & 01 & +0.01 & $\pm$0.01\\
\raisebox{1.5ex}[0pt]{$\log g$} &
\multicolumn{2}{c}{\raisebox{1.5ex}[0pt]{+0.02~dex}} & 63\,077 &
+0 & 00 & +0 & 00 & +0 & 00 & +0 & 01 & +0 & 00 & +0 & 00 & +0 &
00 &
+0 & 01 & +0 & 01 & +0 & 01 & +0.00 & $\pm$0.01\\
& & & 59\,984 & +0 & 00 & +0 & 01 & +0 & 00 & +0 & 00 & +0 & 00 &
+0 & 02 & +0 & 00 &
+0 & 01 & +0 & 01 & +0 & 01 & +0.01 & $\pm$0.01\\
\hline  & & & 160\,691 & $-$0 & 01 & $-$0 & 02 & +0 & 00 & $-$0 &
01 & $-$0 & 01 & +0 & 00
& $-$0 & 02 & $-$0 & 01 & $-$0 & 01 & $-$0 & 01 & $-$0.01 & $\pm$0.01\\
& \multicolumn{2}{c}{\raisebox{1.5ex}[0pt]{+0.05 km s$^{-1}$}} &
22\,484 & $-$0 & 01 & $-$0 & 01 & $-$0 & 01 & $-$0 & 02 & +0 & 00
& $-$0 & 01
& $-$0 & 01 & $-$0 & 01 & $-$0 & 01 & $-$0 & 01 & $-$0.01 & $\pm$0.00\\
\cline{2-3} \raisebox{1.5ex}[0pt]{$\xi$} & & & 63\,077 & $-$0 & 03
& $-$0 & 03 & $-$0 & 01 & $-$0 & 04 & +0 & 00 & +0 & 00 &
+0 & 00 & $-$0 & 02 & $-$0 & 01 & $-$0 & 01 & $-$0.01 & $\pm$0.01\\
& \multicolumn{2}{c}{\raisebox{1.5ex}[0pt]{+0.23 km s$^{-1}$}} &
59\,984 & $-$0 & 03 & $-$0 & 03 & +0 & 00 & $-$0 & 04 & +0 & 00 &
+0 & 00
& $-$0 & 01 & $-$0 & 01 & $-$0 & 01 & $-$0 & 01 & $-$0.01 & $\pm$0.01\\
\hline & $\mbox{EW}<20\mbox{~m\AA}$: & +15\% & 160\,691 & +0 & 06
& +0 & 07 & +0 & 04 & +0 & 07 & +0 & 06 & +0 & 03 & +0 & 07 &
+0 & 04 & +0 & 05 & +0 & 04 & +0.05 & $\pm$0.01\\
\raisebox{1ex}[0pt]{EW} & $\mbox{EW}>20\mbox{~m\AA}$: & +5\% &
22\,484 & +0 & 06 & +0 & 06 & +0 & 08 & +0 & 07 & +0 & 03 & +0 &
06 & +0 & 06 &
+0 & 06 & +0 & 08 & +0 & 08 & +0.06 & $\pm$0.02\\
\cline{2-3} \raisebox{1.5ex}[0pt]{and} &
$\mbox{EW}<15\mbox{~m\AA}$: & +19\% & 63\,077 & +0 & 07 & +0 & 05
& +0 & 09 & +0 & 08 & +0 & 06 & +0 & 09 & +0 & 10 &
+0 & 10 & +0 & 10 & +0 & 10 & +0.08 & $\pm$0.02\\
\raisebox{2ex}[0pt]{$\log gf$} & $\mbox{EW}>15\mbox{~m\AA}$: &
+5\% & 59\,984 & +0 & 06 & +0 & 06 & +0 & 10 & +0 & 06 & +0 & 05 &
+0 & 08 & +0 & 08 &
+0 & 09 & +0 & 10 & +0 & 09 & +0.08 & $\pm$0.02\\
\hline & & & 160\,691 & +0 & 01 & +0 & 01 & +0 & 01 & +0 & 01 & +0
& 00 & +0 & 01 & +0 & 01 &
+0 & 03 & +0 & 04 & +0 & 04 & +0.02 & $\pm$0.01\\
 & \multicolumn{2}{c}{\raisebox{1.5ex}[0pt]{Fe: +0.07~dex}} &
 22\,484 & +0 & 00 & +0 & 01 & +0 & 00 & +0 & 01 & +0 & 00 & +0 &
 01
& +0 & 01 &
+0 & 03 & +0 & 03 & +0 & 03 & +0.01 & $\pm$0.01\\
\raisebox{1.5ex}[0pt]{[Fe/H]} &
\multicolumn{2}{c}{\raisebox{0.5ex}[0pt]{Other elements:}}
 & 63\,077 & +0
& 00 & +0 & 01 & +0 & 01 & +0 & 01 & +0 & 00 & +0 & 00 & +0 & 00 &
+0 & 02 & +0 & 03 & +0 & 01 & +0.01 & $\pm$0.01\\
& \multicolumn{2}{c}{\raisebox{0.5ex}[0pt]{+0.10~dex}} & 59\,984 &
+0 & 00 & +0 & 01 & +0 & 00 & +0 & 01 & +0 & 01 & +0 & 01 & +0 &
01 &
+0 & 02 & +0 & 02 & +0 & 01 & +0.01 & $\pm$0.01\\
\hline \hline & & & 160\,691 & 0 & 09 & 0 & 09 & 0 & 12 & 0 & 10 &
0 & 05 & 0 & 09 & 0 & 09 & 0 & 09 &
0 & 12 & 0 & 12 & 0.08 & $\pm$0.02\\
\raisebox{-0.3ex}[0pt]{Total} & & & 22\,484 & 0 & 09 & 0 & 10 & 0
& 06 & 0 & 10 & 0 & 09 & 0 & 05 & 0 & 10 & 0 & 07 &
0 & 08 & 0 & 07 & 0.10 & $\pm$0.02\\
\raisebox{0.3ex}[0pt]{uncertainty} & & & 63\,077 & 0 & 09 & 0 & 09
& 0 & 14 & 0 & 10 & 0 & 08 & 0 & 12 & 0 & 12 & 0 & 13 &
0 & 14 & 0 & 13 & 0.12 & $\pm$0.02\\
 & & & 59\,984 & 0 & 11 & 0 & 08 & 0 & 13 & 0 & 12 & 0 & 09 & 0 &
13 & 0 & 14 & 0 & 14 &
0 & 15 & 0 & 14 & 0.11 & $\pm$0.02\\
\hline \hline \raisebox{-0.3ex}[0pt]{Average} & & & & 0 & 10 & 0 &
09 & 0 & 11 & 0 & 11 & 0 & 08
  & 0 & 10  & 0 & 11  & 0 & 11  & 0 & 12  &
 0 & 12  &  & \\
\raisebox{0.3ex}[0pt]{total} & & & & $\pm$0 & 01 & $\pm$0 & 01 &
$\pm$0 & 04 & $\pm$0 & 01 & $\pm$0 & 02 &
 $\pm$0 & 04 & $\pm$0 & 02 & $\pm$0 & 03 & $\pm$0 & 03 & $\pm$0 & 03 &
 \raisebox{1.5ex}[0pt]{0.10} & \raisebox{1.5ex}[0pt]{$\pm$0.03}\\
\hline
\end{tabular}

\vspace{0.1cm}

Note: Contributions from EW and $\log gf$ are equal. The values
which are presented correspond to \emph{each} contribution, and
not to the total of these two sources.

Therefore, they are taken in consideration \emph{twice} when
calculating the total uncertainties. \vfill\end{minipage}
\end{sidewaystable*}

\subsection{Adopted abundances}

\begin{figure*}
\centering
\includegraphics*[width=17cm]{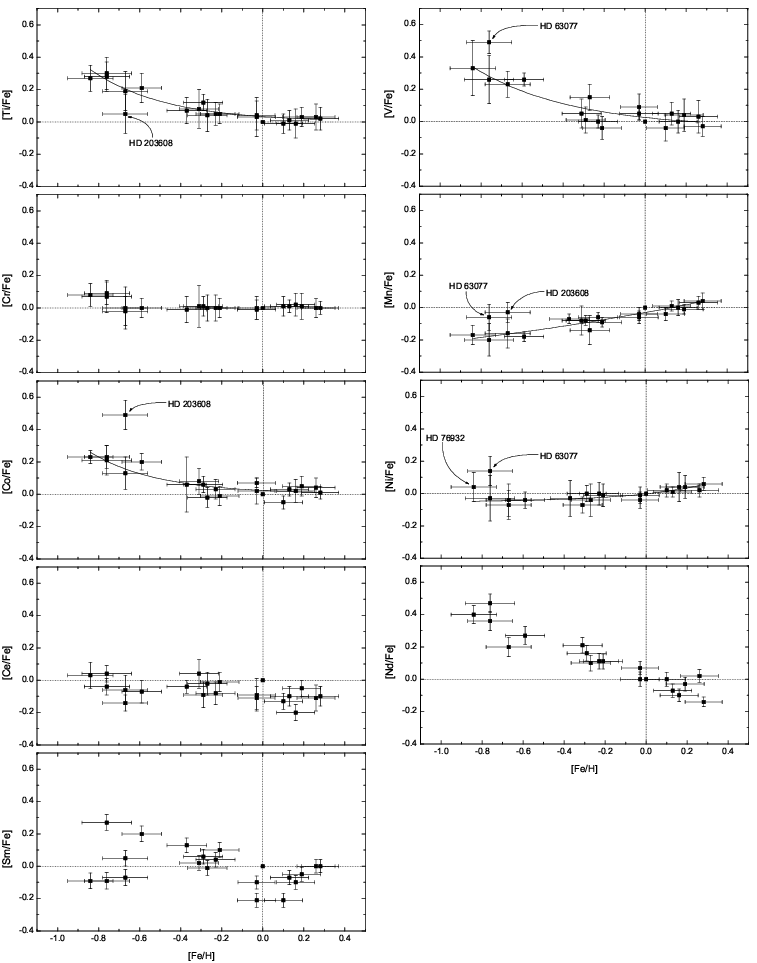}
\caption{Abundance patterns for all contaminating elements.
Outliers are indicated by arrows labelled with their respective HD
numbers. Solid lines, when present, represent exponential fits to
the well behaved data points.} \label{fig:abundances}
\end{figure*}

Figure~\ref{fig:abundances} presents the abundance patterns for
all contaminating elements. Error bars were taken as the
dispersion between the lines or the estimated [element/Fe]
uncertainty, whichever is the largest (see
Tables~\ref{tab:abundances_1} and \ref{tab:abundances_2}). The
presence of a few ($\sim4\%$) outliers can be noticed. For these
stars, these abundances will not be used for the Eu and Th
spectral syntheses. In these cases, the discrepant abundances will
be substituted by the value they would have if they agreed with an
exponential fit to the well behaved data points. One might wonder
if the discrepant abundances would be a hint of chemical
peculiarity. If this was true, one such star would exhibit
peculiar abundances for all elements produced by one specific
nucleosynthesis process, and this behaviour is not observed. Mark
that the three stars which present discrepant abundances
(\object{HD~63\,077}, \object{HD~76\,932}, and
\object{HD~203\,608}) are all metal-poor, with
$\mbox{[Fe/H]}<-0.67$, and therefore more sensitive to noise and
continuum placement errors. The Nd and V abundances of
\object{HD~196\,378} and \object{HD~203\,608}, which have no
measured lines, were established this same way.

\section{Eu and Th abundance determination}

The Eu and Th abundance determination was carried out, for our
sample stars, using a single absorption line for each element. The
only acceptably strong Th line, which allows an accurate abundance
determination, is the \ion{Th}{ii} line located at 4019.13~\AA.
This line is, nevertheless, blended with many others, rendering
imperative the use of spectral synthesis. One other line is often
cited in the literature: a \ion{Th}{ii} line located at
4086.52~\AA. Unfortunately, although less contaminated than the
line at 4019.13~\AA, it is too weak to be used.

Eu has only one adequately strong and uncontaminated line, located
at 4129.72~\AA, available for abundance determinations. However,
this line has a significantly non-Gaussian profile by virtue of
its HFS and isotope shift. Therefore, spectral synthesis is also
required for this element. In stars of other spectral types and
other luminosity classes (e.g. giants), it is possible to use
other Th and Eu lines. This is, nonetheless, not true for our
sample stars (late-F and G dwarfs/subgiants).

The code used to calculate the synthetic spectra was kindly made
available to us by its developer, Monique Spite. This code
calculates a synthetic spectrum based on a list of lines with
their respective atomic parameters ($\log gf$, and excitation
potential of the lower level of the transition $\chi$, in eV). We
employed the same model atmospheres and partition functions used
in the determination of atmospheric parameters and abundances of
contaminating elements (Section~\ref{sec:model_atm_part_funct}).
The only exception was for Th, whose partition functions were
calculated by us, based on data from H. Holweger that was
published as a \emph{private communication} in MKB92. Irwin's data
are too inaccurate, because for the singly and doubly ionised
states of Th, he scales the partition functions along
iso-electronic sequences from a lower mass element, due to lack of
specific Th data. This resulted in values much lower than those of
Holweger, which were calculated by summing a large number of Th
atomic levels.

\subsection{Europium}

\subsubsection{Spectral synthesis}
\label{sec:eu_abundance_determination}

The list of lines used for calculation of the synthetic spectrum
of the Eu region was based on the line lists of the Vienna Atomic
Line Database -- VALD \citep{kupkaetal99}. Initially, we took all
lines found between 4128.4~\AA\ and 4130.4~\AA. The first cut
eliminated all lines of elements ionised two or more times, and
all lines with lower-level excitation potential greater than
10~eV, since these lines are unobservable in the photospheres of
stars with effective temperatures as low as those in our sample.
As a next step, we calculated synthetic spectra for the four stars
with the most extreme effective temperatures and metallicities in
our sample (\object{HD~22\,484}, \object{HD~59\,984},
\object{HD~63\,077}, and \object{HD~203\,608}) using laboratory
$\log gf$ values, and the atmospheric parameters and abundances
previously obtained. Lines with EWs lower than 0.01~m\AA\ in these
four stars were removed. The \ion{Sc}{i} line located at
4129.750~\AA, cited as important by \citet{mashonkina&gehren00},
presents negligible EW in all four standard stars. 23 Ti, V, Cr,
Fe, Co, Nb, Ce, Pr, Nd, Sm, and Dy lines were kept, besides the Eu
line itself. Adopted wavelengths were taken from the VALD list
\citep{bard&kock94,kurucz93,kurucz94,whaling83,wickliffeetal94}.
To improve the fits further, three artificial \ion{Fe}{i} lines
were added as substitutes for unknown blends
\citep{lawleretal01,mashonkina&gehren00}; the influence of these
artificial lines in the obtained Eu abundances is nevertheless
small, as they are located relatively far from the center of the
\ion{Eu}{ii} line. The final list is presented in
Table~\ref{tab:eu_line_list}.

In the spectral synthesis calculations, the \ion{Eu}{ii} line was
substituted by its HFS components, calculated by us using
Casimir's equation \citep{casimir_63}, and HFS laboratory data
from \citet{beckeretal93}, \citet{brostrometal95},
\citet{villemoes&wang94}, and \citet{molleretal93}. Isotope shift
was taken into account, using data from \citet{brostrometal95} and
the solar abundance isotopic ratio
$\varepsilon(\mbox{\element[][151]{Eu}})/\varepsilon(\mbox{\element[][153]{Eu}})=1.00\pm0.29$
\citep{lawleretal01}. The solar value was used for all stars
because there are strong indications that it is not
metallicity-dependent, since several works dealing with r-process
nucleosynthesis in very-metal-poor halo stars find similar results
\citep{pfeifferetal97,cowanetal99,snedenetal02,aokietal03}.

\begin{table*}
\caption[]{Line list used in the spectral synthesis of the
\ion{Eu}{ii} line at 4129.72~\AA.} \label{tab:eu_line_list}
\begin{tabular}{ c c c c c c c }
\hline \hline $\lambda$ (\AA) & Element & $\chi$ (eV) & $\log
gf_{\mathrm{CES}}$
& $\log gf_{\mathrm{FEROS}}$ & $\log gf$ source & Obs.\\
 \hline
4128.742  &  \ion{Fe}{ii} & 2.580  & $-$3.832 & $-$3.554 & solar fit & \\
4129.000  &  \ion{Cr}{i}  & 4.210  & $-$2.603 & $-$2.603 & VALD & \\
4129.040  &  \ion{Ru}{i}  & 1.730  & $-$1.030 & $-$1.030 & VALD & \\
4129.147  &  \ion{Pr}{ii} & 1.040  & $-$0.100 & $-$0.100 & VALD & \\
4129.159  &  \ion{Ti}{ii} & 1.890  & $-$2.330 & $-$2.210 & solar fit & \\
4129.166  &  \ion{Ti}{i}  & 2.320  &   +0.131 &   +0.251 & solar fit & \\
4129.165  &  \ion{Cr}{i}  & 3.010  & $-$1.948 & $-$1.948 & VALD & \\
4129.174  &  \ion{Ce}{ii} & 0.740  & $-$0.901 & $-$0.901 & VALD & \\
4129.196  &  \ion{Cr}{i}  & 2.910  & $-$1.374 & $-$1.254 & solar fit & \\
4129.220  &  \ion{Fe}{i}  & 3.420  & $-$2.280 & $-$2.160 & solar fit & \\
4129.220  &  \ion{Sm}{ii} & 0.250  & $-$1.123 & $-$1.123 & VALD & \\
4129.425  &  \ion{Dy}{ii} & 0.540  & $-$0.522 & $-$0.522 & VALD & \\
4129.426  &  \ion{Nb}{i}  & 0.090  & $-$0.780 & $-$0.780 & VALD & \\
4129.461  &  \ion{Fe}{i}  & 3.400  & $-$2.180 & $-$1.920 & solar fit & \\
4129.530  &  \ion{Fe}{i}  & 3.140  & $-$3.425 & $-$3.455 & solar fit & artificial\\
4129.610  &  \ion{Fe}{i}  & 3.500  & $-$3.700 & $-$3.730 & solar fit & artificial\\
4129.643  &  \ion{Ti}{i}  & 2.240  & $-$1.987 & $-$1.987 & VALD & \\
4129.657  &  \ion{Ti}{i}  & 2.780  & $-$2.297 & $-$2.297 & VALD & \\
4129.721  &  \ion{Eu}{ii} & 0.000  &   +0.173 &   +0.173 & \citet{komarovskii91} & HFS\\
4129.817  &  \ion{Co}{i}  & 3.810  & $-$1.808 & $-$1.808 & VALD & \\
4129.837  &  \ion{Nd}{ii} & 2.020  & $-$0.543 & $-$0.543 & VALD & \\
4129.965  &  \ion{Fe}{i}  & 2.670  & $-$3.390 & $-$3.290 & solar fit & artificial\\
4129.994  &  \ion{V}{i}   & 2.260  & $-$1.769 & $-$1.769 & VALD & \\
4130.037  &  \ion{Fe}{i}  & 1.560  & $-$4.195 & $-$4.030 & solar fit & \\
4130.038  &  \ion{Fe}{i}  & 3.110  & $-$2.385 & $-$2.225 & solar fit & \\
4130.068  &  \ion{Fe}{i}  & 3.700  & $-$3.763 & $-$3.763 & VALD & \\
4130.073  &  \ion{Cr}{i}  & 2.910  & $-$1.971 & $-$1.971 & VALD & \\
\hline
\end{tabular}

References: see text.
\end{table*}

\begin{figure}
\resizebox{\hsize}{!}{\includegraphics*{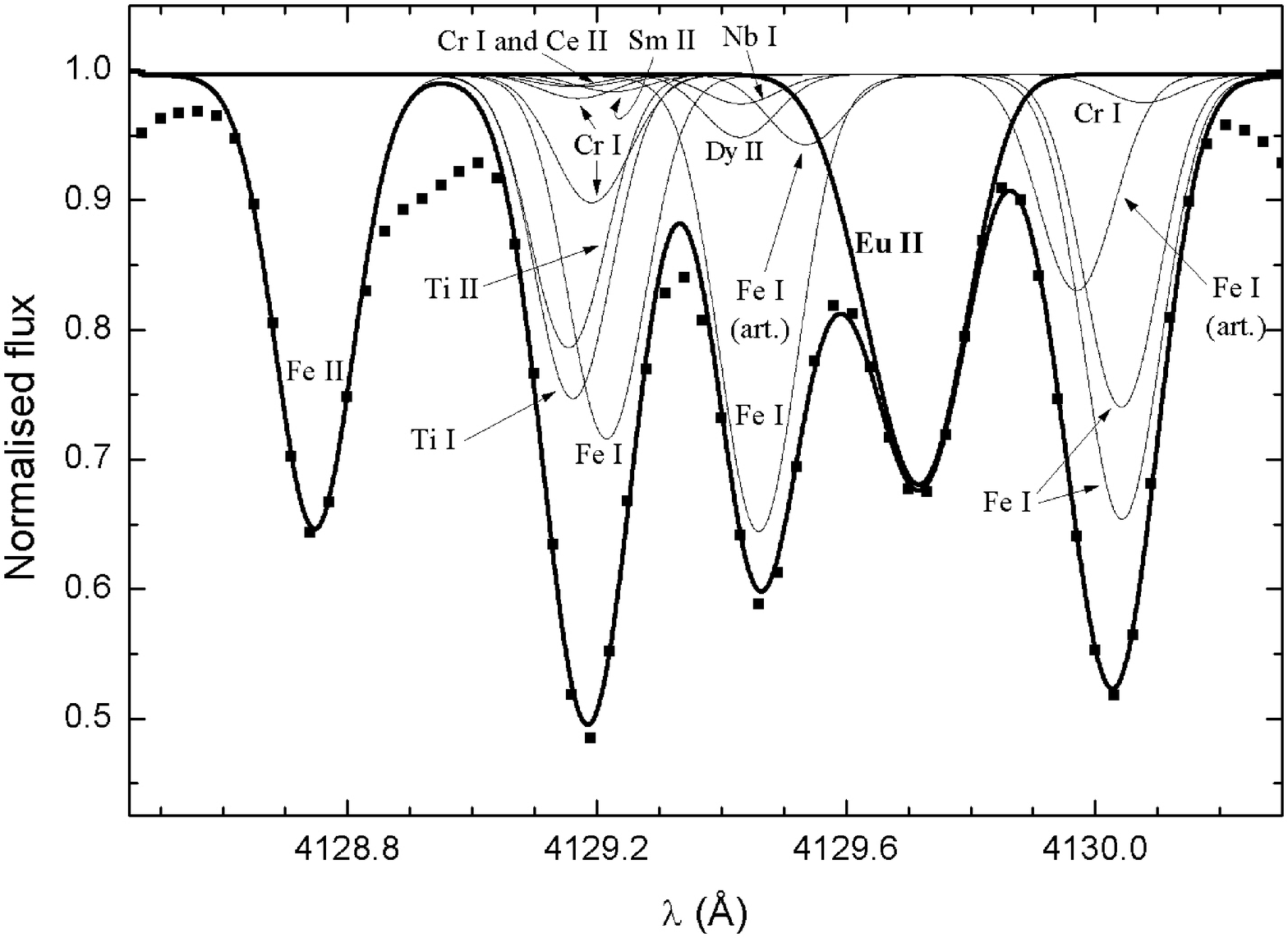}}
\caption{Synthesized spectral region for the \ion{Eu}{ii} line at
4129.72~\AA, for the Sun. The two thick lines represent the total
synthetic spectrum and the \ion{Eu}{ii} line. The thin lines
represent the most dominant lines in the region, including two
artificial \ion{Fe}{i} lines. Points are the observed FEROS solar
spectrum.} \label{fig:eu_sun}
\end{figure}

We adopted a projected rotation velocity for the integrated solar
disk $v\,\sin i=1.8~\mbox{km s}^{-1}$
\citep{mashonkina00,mashonkina&gehren00}. The width of the
Gaussian profile convolved to the synthetic spectrum to take
macroturbulence and instrumental broadenings into account was
obtained by fitting the \ion{Fe}{ii} located at 4128.742~\AA,
which is sufficiently isolated for this purpose
\citep{woolfetal95,koch&edvardsson02}. Solar $\log gf$ were
obtained by keeping abundances fixed at their solar values
\citep{grevesse&sauval98}, and fitting the observed solar
spectrum. Only the stronger Ti, Cr, and Fe lines had their $\log
gf$ adjusted. For the \ion{Eu}{ii} line, we adopted a fixed $\log
gf$ from \citet{komarovskii91}; this value was distributed among
the HFS components according to relative intensities calculated by
White \& Eliason, tabulated in \citet{condon&shortley67}. The
other, weaker lines had their $\log gf$ kept at their laboratory
values adopted from the VALD list
\citep{kurucz93,kurucz94,wickliffeetal94}. As we kept the Eu $\log
gf$ fixed, we allowed its abundance to vary. The complete solar
spectrum fit procedure was carried out iteratively because it
requires alternate Gaussian profile adjustments (which modify the
\emph{shape} of the lines) and $\log gf$ adjustments (which modify
the \emph{EWs} of the lines). The solar $\log gf$ and Gaussian
profile determination was accomplished independently for the two
sources of Eu spectra used, leading to different Eu solar
abundances ($\log\varepsilon(\mbox{Eu})_{\sun}=+0.37$ and $+0.47$
for the CES and FEROS, respectively). The stellar [Eu/H] abundance
ratios, used in this work in lieu of the absolute abundances, were
obtained by subtracting the appropriate solar value from the
stellar absolute abundances
($\mbox{[Eu/H]}=\log\varepsilon(\mbox{Eu})_{\mathrm{star}}-
\log\varepsilon(\mbox{Eu})_{\sun}$). Figure~\ref{fig:eu_sun}
presents the observed and synthetic solar spectrum in the Eu
region. The strongest lines that compose the total synthetic
spectrum are presented independently.

For the Eu synthesis in the sample stars, we kept the abundances
of all elements other than Eu fixed at the values determined using
EWs, in Sect.~\ref{sec:chemical_abundances}. Small adjustments
were allowed only to the abundances of Ti, Cr, and Fe, in order to
improve the fit. The adjustments were kept within the
uncertainties of the abundances of these elements -- see
Sect.~\ref{sec:chemical_abundances}. The abundances of elements
for which we did not measure any EW were determined scaling the
solar abundances, following elements produced by the same
nucleosynthetic processes. We chose to scale the Sm abundances as
well, even though they were determined by us, because our results
for this element exhibit high scatter and absence of well-defined
behaviour (see Fig.~\ref{fig:abundances}). Sm and Dy abundances
were obtained following Eu (produced mainly by r-process,
according to \citealt{burrisetal00}); Nb, Rb, and Pr were obtained
following Nd (produced by both r-process and s-process, also
according to \citealt{burrisetal00}). Mark that the influence of
these lines is marginal, since they are weak and are not close to
the Eu line. Projected rotation velocities $v\,\sin i$ and
macroturbulence and instrumental broadening Gaussian profiles were
determined by fitting the \ion{Fe}{ii} line at 4128.742~\AA, as
done with the solar spectrum.

\begin{figure*}
\centering
\includegraphics*[width=17cm]{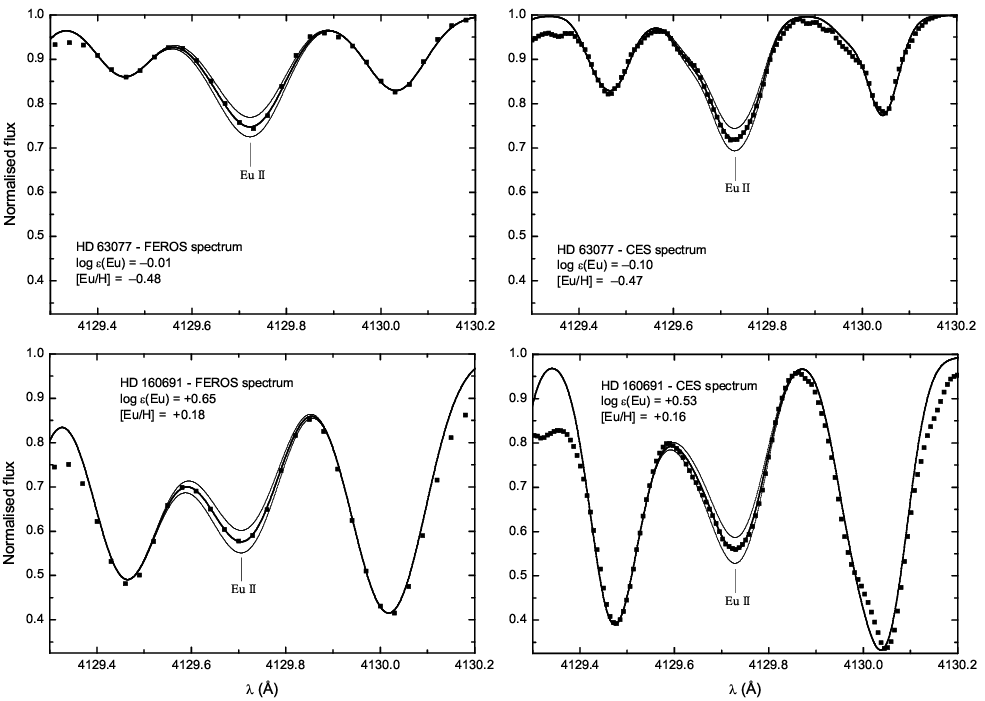}
\caption{Examples of spectral syntheses of the \ion{Eu}{ii} line
at 4129.72~\AA. FEROS and CES spectra are presented for two stars
with extreme Eu abundances (\object{HD~63\,077} and
\object{HD~160\,691}). Points are the observed spectra. Thick
lines are the best fitting synthetic spectra, calculated with the
shown Eu abundances. Thin lines represent variations in the Eu
abundance $\Delta\log\varepsilon(\mbox{Eu})=\pm0.05~\mbox{dex}$.
The shown abundances are the ones that best fit the presented
spectra, and not the average values that are presented in
Table~\ref{tab:th_eu_abundances}.} \label{fig:examples_eu}
\end{figure*}

Stars observed more than once had their spectra analysed
independently, and the results were averaged. Abundances obtained
with the different spectra of one object presented a maximum
variation of 0.02~dex. Examples of spectral syntheses can be seen
in Fig.~\ref{fig:examples_eu} for the two stars with extreme Eu
abundances that were observed with both FEROS and CES
(\object{HD~63\,077} and \object{HD~160\,691}). Note the effect of
CES spectra higher sampling, which better puts the asymmetric
profile of the \ion{Eu}{ii} line in evidence.

\begin{figure}
\resizebox{\hsize}{!}{\includegraphics*{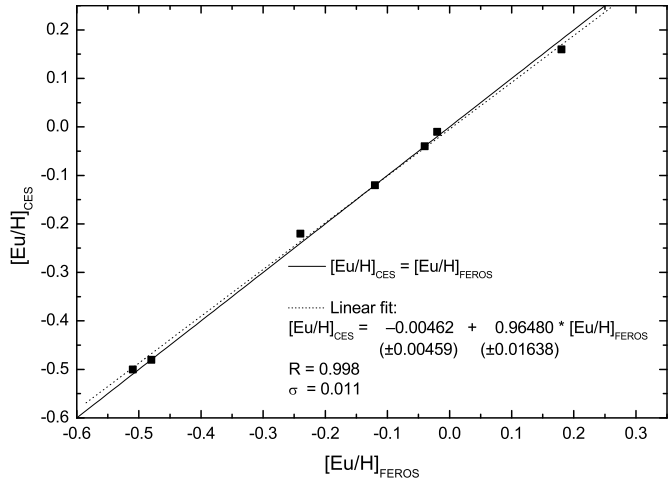}}
\caption{Comparison between Eu abundances determined using CES and
FEROS spectra. Solid line is the identity. Dotted line is a linear
fit, whose parameters are shown in the figure.}
\label{fig:eu_ces_eu_feros}
\end{figure}

\begin{figure}
\resizebox{\hsize}{!}{\includegraphics*{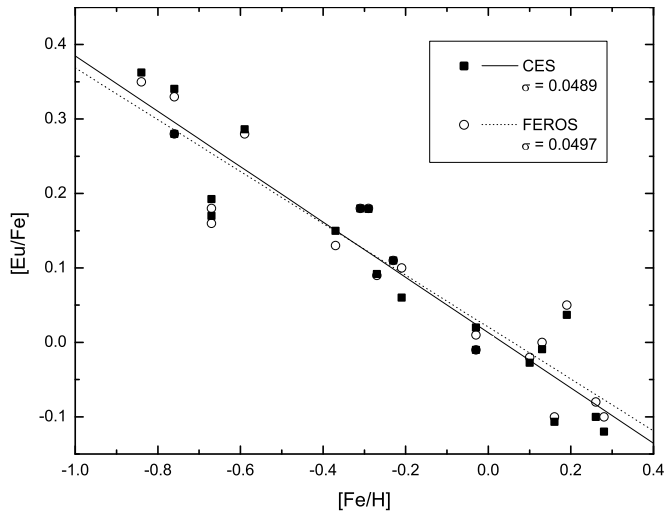}}
\caption{[Eu/Fe] vs. [Fe/H] diagram for all sample stars.
Abundances obtained with CES and FEROS spectra are presented. The
Sun is not included in the comparison, since its abundance is
always equal to zero by definition. Solid and dotted lines are
linear fits to these two data sets, whose scatters are shown in
the figure. Mark that data from the CES are slightly better than
those from FEROS.} \label{fig:eu_fe_fe_h_ces_feros}
\end{figure}

Eu abundances obtained using CES and FEROS spectra are compared in
Fig.~\ref{fig:eu_ces_eu_feros}. A linear fit was calculated, and
the resulting relation was used to convert FEROS results into the
CES system. This way, we ended up with two sets of Eu abundances:
one totally homogeneous, obtained with FEROS spectra, and one
obtained partially with CES spectra and partially with
corrected-into-CES-system FEROS spectra. We plotted a [Eu/Fe] vs.
[Fe/H] diagram for both sets of abundances, presented in
Fig.~\ref{fig:eu_fe_fe_h_ces_feros}. Linear fits calculated for
each set show that the CES abundances present a lower
scatter~$\sigma$, and for this reason the CES results were adopted
for all subsequent analyses.

\subsubsection{Uncertainty assessment}

[Eu/H] abundance uncertainties were estimated with a procedure
identical to the one used for the abundances of contaminating
elements determined with EWs, thoroughly described in
Sect.~\ref{sec:errors_assessment}. It consists basically of
varying the atmospheric parameters and continuum position
independently, by an amount equal to their uncertainties, and
recalculating the synthetic spectra. Total uncertainties are
obtained by RMS of the individual sources of uncertainty. We chose
four stars as standards for the assessment, with extreme effective
temperatures and metallicities: \object{HD~160\,691} (cool and
metal-rich), \object{HD~22\,484} (hot and metal-rich),
\object{HD~63\,077} (cool and metal-poor), and
\object{HD~203\,608} (hot and metal-poor) --
Fig.~\ref{fig:error_stars}.

Continuum placement is the preponderant source of Eu abundance
uncertainty. The placement uncertainty itself is higher for
metal-rich stars, as these present stronger absorption lines,
lowering the apparent continuum which restricts the choice of good
normalisation windows. To estimate the influence of continuum
placement uncertainties, we multiplied the spectral flux of the
metal-rich standard stars  by 0.98, and that of the metal-poor
standard stars by 0.99, and recalculated the synthetic spectra.

Table~\ref{tab:th_eu_uncertainties} presents the [Eu/H], [Eu/Fe],
[Th/H], [Th/Fe], and [Th/Eu] uncertainties for each standard star.
Metallicity and continuum placement clearly dominate.
Microturbulence velocity variations have no effect in the [Eu/H]
abundances. This happens because the HFS broadens the line so much
that it becomes nearly unsaturated, as found by \citet[\space
WTL95]{woolfetal95}, \citet{lawleretal01} and \citet[\space
KE02]{koch&edvardsson02}. [Element/Fe] uncertainties were obtained
simply by subtracting the [Fe/H] uncertainty from the [element/H]
value (e.g.,
$\mbox{uncert.}_{\mathrm{[Eu/Fe]}}=\mbox{uncert.}_{\mathrm{[Eu/H]}}-\mbox{uncert.}_{\mathrm{[Fe/H]}}$).
Total uncertainties for the other sample stars were obtained by
weighted average of the standard stars values, using as weight the
reciprocal of the distance of the star to each standard star, in
the $T_{\mathrm{eff}}$ vs. [Fe/H] plane; this procedure is
identical to the one used for the uncertainties of contaminating
elements (Sect.~\ref{sec:errors_assessment}).
Table~\ref{tab:th_eu_abundances} presents the final [Eu/H],
[Th/H], and [Th/Eu] abundance ratios, for all sample stars, along
with their respective uncertainties relative to H and Fe.

Non-LTE effects in the abundances obtained with the \ion{Eu}{ii}
line at 4129.72~\AA\ have been calculated by
\citet{mashonkina&gehren00}. In the line formation layers, the
ground state was found to be slightly underpopulated, and the
excited level overpopulated. As a consequence, our results would
show a small difference if our method employed absolute
abundances. But since we employ a differential analysis, these
differences cancel each other partially, becoming negligible
(WTL95 and KE02).

\subsubsection{Comparison with literature
results}

We conducted a painstaking search of the literature for works with
Eu abundances, and selected WTL95 and KE02 as the ones with the
most careful determinations of disk stars, as well as a sizable
sample of high statistical significance. Their sample is composed
of a subset from \citet{edvardssonetal93}, and they use the same
atmospheric parameters (obtained from Str\"omgren photometric
calibrations). The Eu abundances were determined by spectral
synthesis using the same line we used, with a procedure
fundamentally identical to ours. KE02 merged their database with
that from WTL95 by means of a simple linear conversion, obtained
by intercomparison. We used this merged set of abundances to
compare to our results.

\begin{figure}
\resizebox{\hsize}{!}{\includegraphics*{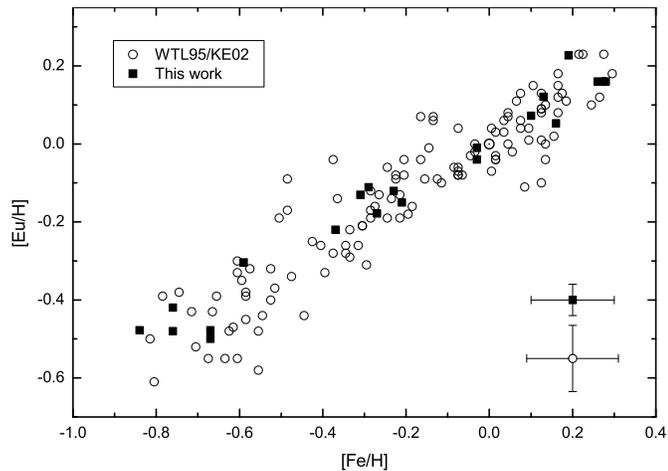}}
\caption{[Eu/H] vs. [Fe/H] diagram for our sample stars and those
from WTL95 and KE02. Average error bars for the two data sets are
provided in the lower right corner. Note that the behaviour of
both data sets is similar, but that our abundances present
considerably lower scatter.} \label{fig:eu_h_fe_h_wtl95_ke02}
\end{figure}

Figure~\ref{fig:eu_h_fe_h_wtl95_ke02} presents a [Eu/H] vs. [Fe/H]
diagram with our results and those from WTL95 and KE02. Both data
sets exhibit the same behaviour, but our results present scatter
36\%~lower than those of WTL95 and KE02, even though our sample is
6~times smaller than theirs (linear fits result in
$\sigma_{\mathrm{our~work}}=0.050$ and
$\sigma_{\mathrm{WTL95/KE02}}=0.078$). The lower scatter of our
results is a consequence of improvements introduced in our
analysis, like the use of atmospheric parameters obtained by us
through the detailed and totally self-consistent spectroscopic
analysis carried out in Sect.~\ref{sec:atmospheric_parameters}.
WTL95 and KE02 demonstrated, using Monte Carlo simulations of
observational errors, that the observed scatter is not real, but
mainly a result of observational, analytical, and systematic
errors.

\subsection{Thorium}

\subsubsection{Spectral synthesis}
\label{sec:th_abundance_determination}

The list of lines used for calculation of the synthetic spectrum
of the Th region was constructed following the same procedure used
for the Eu analysis. 14 V, Mn, Fe, Co, Ni, W, Ce, and U lines were
kept, besides the Th line itself. Adopted wavelengths were taken
from the VALD list
\citep{bard&kock94,kurucz93,kurucz94,whaling83}, with the
exception of the \ion{Fe}{i} line at 4019.043~\AA, \ion{Ni}{i} at
4019.067~\AA, and \ion{Th}{ii} at 4019.130~\AA, determined in
laboratory by \citet{learneretal91} with a precision higher than
0.0005~\AA. The \ion{Co}{i} lines were substituted by their HFSs
components, calculated by us using Casimir's equation, and HFS
laboratory data from \citet{childs&goodman68},
\citet{guthohrlein&keller90}, \citet{pickering&thorne96}, and
\citet{pickering96}. To improve the fits further, one artificial
\ion{Fe}{ii} line was added as substitute for unknown blends
(MKB92; \citealt{francoisetal93});  the influence of this
artificial line in the obtained Th abundances is nevertheless
small, as it is located relatively far from the center of the
\ion{Th}{ii} line. The final list is presented in
Table~\ref{tab:th_line_list}.

\begin{table*}
\caption[]{Line list used in the spectral synthesis of the
\ion{Th}{ii} line at 4019.13~\AA.} \label{tab:th_line_list}
\begin{tabular}{ c c c c c c c }
\hline \hline $\lambda$ (\AA) & Element & $\chi$ (eV) & $\log
gf_{\mathrm{3.60~m}}$
& $\log gf_{\mathrm{CAT}}$ & $\log gf$ source & Obs.\\
 \hline
4018.986 & \ion{U}{ii} & 0.04 & $-$1.391 & $-$1.391 & VALD & \\
4018.999 & \ion{Mn}{i}  & 4.35 & $-$1.497 & $-$1.497 & VALD & \\
4019.036 & \ion{V}{ii} & 3.75 & $-$2.704 & $-$2.704 & VALD & \\
4019.042 & \ion{Mn}{i}  & 4.67 & $-$1.031 & $-$1.026 & solar fit & \\
4019.043 & \ion{Fe}{i}  & 2.61 & $-$3.150 & $-$3.145 & solar fit & \\
4019.057 & \ion{Ce}{ii} & 1.01 & $-$0.470 & $-$0.445 & solar fit & \\
4019.067 & \ion{Ni}{i}  & 1.94 & $-$3.404 & $-$3.329 & solar fit & \\
4019.130 & \ion{Th}{ii} & 0.00 & $-$0.228 & $-$0.228 & \citet{nilssonetal02} & \\
4019.132 & \ion{Co}{i}  & 2.28 & $-$2.270 & $-$2.270 & \citet{lawleretal90} & HFS\\
4019.134 & \ion{V}{i}  & 1.80 & $-$1.300 & $-$1.300 & VALD & \\
4019.206 & \ion{Fe}{ii} & 3.00 & $-$5.380 & $-$5.425 & solar fit & artificial\\
4019.228 & \ion{W}{i}  & 0.41 & $-$2.200 & $-$2.200 & VALD & \\
4019.293 & \ion{Co}{i}  & 0.58 & $-$3.232 & $-$3.232 & VALD & HFS\\
4019.297 & \ion{Co}{i}  & 0.63 & $-$3.769 & $-$3.769 & VALD & HFS\\
\hline
\end{tabular}

References: see text.
\end{table*}

\begin{figure}
\resizebox{\hsize}{!}{\includegraphics*{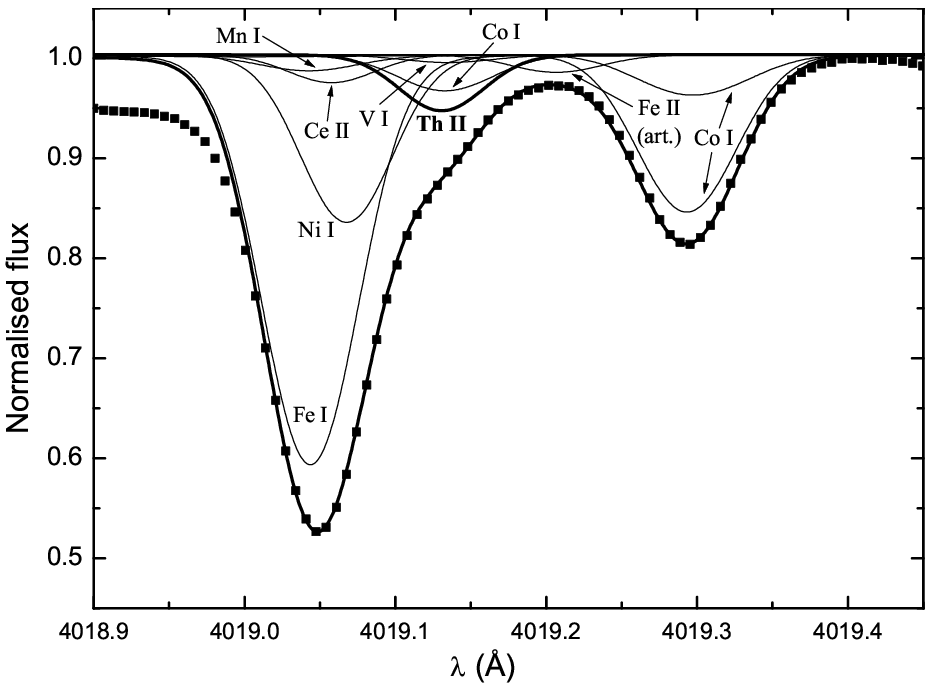}}
\caption{Synthesized spectral region for the \ion{Th}{ii} line at
4019.13~\AA, for the Sun. The two thick lines represent the total
synthetic spectrum and the \ion{Th}{ii} line. The thin lines
represent the most dominant lines in the region, including an
artificial \ion{Fe}{ii} line. Points are the observed CES solar
spectrum, obtained with the 3.60~m telescope.} \label{fig:th_sun}
\end{figure}

The width of the Gaussian profile used to take macroturbulence and
instrumental broadenings into account was obtained by fitting the
\ion{Co}{i} lines located at 4019.293~\AA\ and 4019.297~\AA. Solar
$\log gf$ were obtained by keeping abundances fixed at their solar
values \citep{grevesse&sauval98}, and fitting the observed solar
spectrum. Only the stronger Mn, Fe, Ce, and Ni lines had their
$\log gf$ adjusted. For the \ion{Th}{ii} line, we adopted a fixed
$\log gf$ from \citet{nilssonetal02}; for the \ion{Co}{i} lines we
adopted fixed $\log gf$ from \citealt{lawleretal90} (4019.132~\AA)
and VALD \citep[4019.293~\AA\ and 4019.297~\AA,\ ][]{kurucz94}.
The other, weaker lines had their $\log gf$ kept at their
laboratory values adopted from the VALD list
\citep{kurucz93,kurucz94}. As we kept the Th $\log gf$ fixed, we
allowed its abundance to vary. The solar $\log gf$ and Gaussian
profile determination was accomplished independently for each
telescope used, leading to slightly different Th solar abundances
($\log\varepsilon(\mbox{Th})_{\sun}=+0.03$ and $+0.04$ for the
3.60~m and CAT, respectively). As with Eu, the stellar [Th/H]
abundance ratios were obtained by subtracting the appropriate
solar value from the stellar absolute abundances
($\mbox{[Th/H]}=\log\varepsilon(\mbox{Th})_{\mathrm{star}}-\log\varepsilon(\mbox{Th})_{\sun}$).
Figure~\ref{fig:th_sun} presents the observed and synthetic solar
spectrum in the Th region. The strongest lines that compose the
total synthetic spectrum are presented independently.

For the Th synthesis in the sample stars, we kept the abundances
of all elements other than Th fixed at the values determined using
EWs, in Sect.~\ref{sec:chemical_abundances}. Small adjustments
were allowed only to the abundances of Mn, Fe, Ce, and Ni, in
order to improve the fit. The adjustments were kept within the
uncertainties of the abundances of these elements -- see
Sect.~\ref{sec:chemical_abundances}. The W abundances, for which
we did not measure any EW, were determined scaling the solar
abundances, following the Fe abundances. The U abundances were
estimated from a simplifying hypothesis that the [U/H] stellar
abundance at the time of formation is independent of metallicity.
We allied this hypothesis to a simple linear age-metallicity
relation, in which a star with $\mbox{[Fe/H]}=-1.00$ is 10~Gyr
old, and one with $\mbox{[Fe/H]}=+0.00$ is 4.57~Gyr old, i.e.,
$\mbox{age(Gyr)}=4.57-5.43\,\mbox{[Fe/H]}$. Given the known
\element[][238]{U} half-life $t_{1/2}=4.46~\mbox{Gyr}$, we reach a
relation $\mbox{[U/H]}=-0.50+0.37\,\mbox{[Fe/H]}$. Mind that the W
and U lines have only a small influence on the final Th derived by
the analysis, since they are weak (with
$\mbox{EW}<0.2~\mbox{m\AA}$ for the metal-richest stars) and are
not close to the Th line. We employed the projected rotation
velocities $v\,\sin i$ determined in the Eu analysis, and the
macroturbulence and instrumental broadening Gaussian profiles were
obtained by fitting the \ion{Co}{i} lines at 4019.293~\AA\ and
4019.297~\AA, like done with the solar spectrum.

\begin{figure*}
\centering
\includegraphics*[width=17cm]{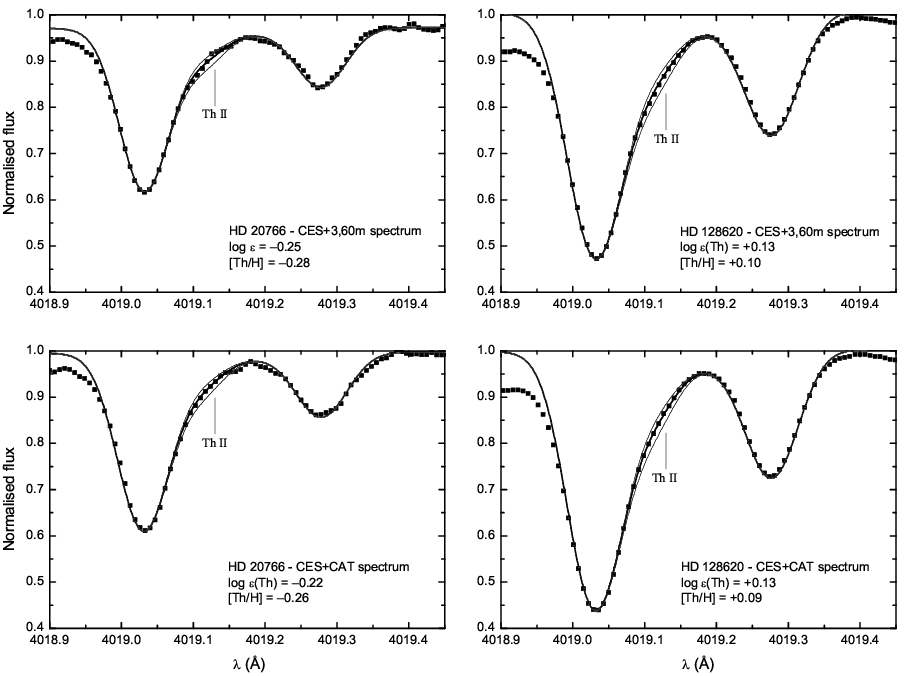}
\caption{Examples of spectral syntheses of the \ion{Th}{ii} line
at 4019.13~\AA. 3.60~m and CAT spectra are presented for two stars
with extreme Th abundances (\object{HD~20\,766} and
\object{HD~128\,620}). Points are the observed spectra. Thick
lines are the best fitting synthetic spectra, calculated with the
shown Th abundances. Thin lines represent variations in the Th
abundance $\Delta\log\varepsilon(\mbox{Th})=\pm0.20~\mbox{dex}$.
The shown abundances are the ones that best fit the presented
spectra, and not the average values that are presented in
Table~\ref{tab:th_eu_abundances}.} \label{fig:examples_th}
\end{figure*}

Stars observed more than once had their spectra analysed
individually, and the results were averaged. Different spectra for
each star presented a 0.02~dex maximum variation. Examples of
spectral syntheses can be seen in Fig.~\ref{fig:examples_th} for
the two stars with extreme Th abundances that were observed with
both 3.60~m and CAT (\object{HD~20\,766} and
\object{HD~128\,620}). Mark that, due to problems that degraded
the resolving power of spectra obtained with the 3.60~m telescope,
these are almost identical to the ones obtained with the CAT.

\begin{figure}
\resizebox{\hsize}{!}{\includegraphics*{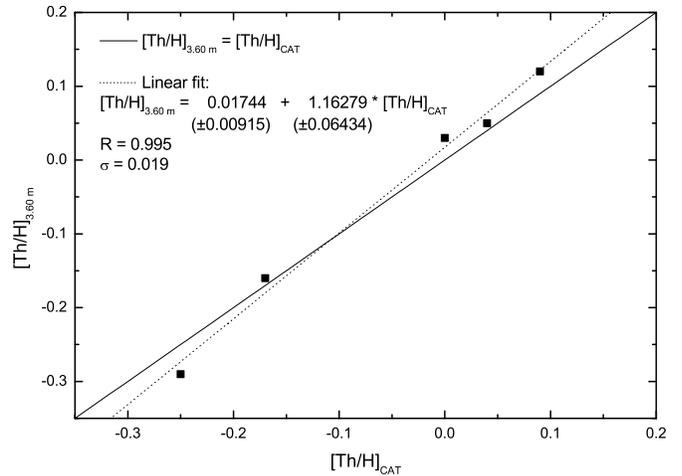}}
\caption{Comparison between Th abundances determined using CES
spectra obtained with the CAT and with the 3.60~m. Solid line is
the identity. Dotted line is a linear fit, whose parameters are
shown in the figure.} \label{fig:th_360_th_cat}
\end{figure}

\begin{figure}
\resizebox{\hsize}{!}{\includegraphics*{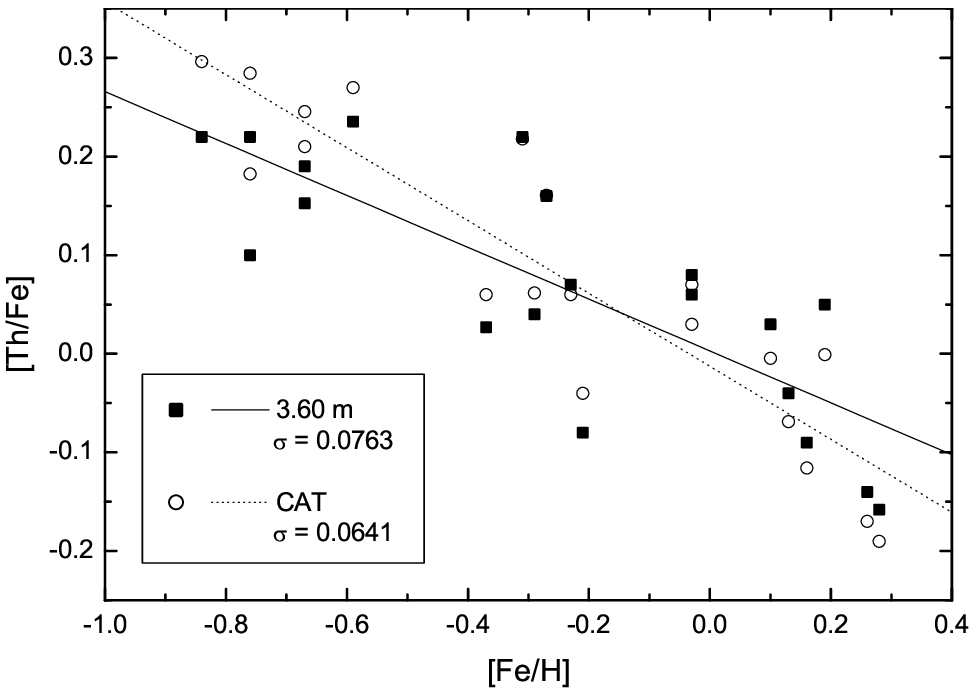}}
\caption{[Th/Fe] vs. [Fe/H] diagram for all sample stars.
Abundances determined using CES spectra obtained with the CAT and
the 3.60~m are presented.  The Sun is not included in the
comparison, since its abundance is always equal to zero by
definition. Solid and dotted lines are linear fits to these two
data sets, whose scatters are shown in the figure. Mark that data
from the CAT are better than those from 3.60~m.}
\label{fig:th_fe_fe_h_360_cat}
\end{figure}

Th abundances determined using spectra obtained with the 3.60~m
and CAT are compared in Fig.~\ref{fig:th_360_th_cat}. A linear fit
was calculated, and the resulting relation was used to convert
3.60~m results into the CAT system, and vice versa. This way, we
ended up with two sets of Th abundances: one obtained partially
with 3.60~m spectra and partially with
corrected-into-3.60-m-system CAT spectra, and one that is just the
opposite, obtained partially with CAT spectra and partially with
corrected-into-CAT-system 3.60~m spectra. We plotted a [Th/Fe] vs.
[Fe/H] diagram for both sets of abundances, presented in
Fig.~\ref{fig:th_fe_fe_h_360_cat}. Linear fits calculated for each
set show that the CAT abundances present a lower scatter~$\sigma$,
and for this reason the CAT results were adopted for all
subsequent analyses. Thus, our adopted Eu and Th were obtained in
one and the same system, related to one single instrument (CES)
and telescope (CAT), reinforcing the homogeneity of the analysis.

\subsubsection{Uncertainty assessment}

[Th/H] abundance uncertainties were obtained with a procedure
identical to the one used for Eu.
Table~\ref{tab:th_eu_uncertainties} contains the results of the Th
uncertainty analysis for the four standard stars, and
Table~\ref{tab:th_eu_abundances} contains the abundances for all
sample stars, with their respective uncertainties. Dependence of
Th abundances on atmospheric parameters is similar to Eu, but
exhibit 3 to 5 times higher sensitivity to continuum placement
variations. This results from the much lower EW of the Th line,
which ranges from 1~m\AA\ to 8~m\AA\ for our sample stars, whereas
Eu ranges from 20~m\AA\ to 90~m\AA. Furthermore, the Eu line is
practically isolated from contaminations, while the Th line is
located in the red wing of a feature composed of strong Mn, Fe,
Ce, and Ni lines, and is blended with one \ion{V}{i} and one
\ion{Co}{i} line that, depending on the metallicity of the star,
has an EW comparable to Th itself.

\begin{table*}
\caption[]{[Eu/H], [Eu/Fe], [Th/H], [Th/Eu], and [Th/Eu] abundance
uncertainties for the standard stars.}
\label{tab:th_eu_uncertainties}
\begin{tabular} { c c c c c c c c }
\hline\hline Parameter & $\Delta$Parameter & HD &
[Eu/H] & [Eu/Fe] & [Th/H] & [Th/Fe] & [Th/Eu]\\
\hline & & 160\,691 & +0.01 & +0.00 & +0.00 & $-$0.01 & $-$0.01\\
& & 22\,484 & +0.01 & $-$0.01 & +0.01 & $-$0.01 & +0.00\\
\raisebox{1.5ex}[0pt]{$T_{\mathrm{eff}}$} &
\raisebox{1.5ex}[0pt]{+27~K} & 63\,077 & +0.01 & $-$0.01 &
 +0.01 & $-$0.01 & +0.00\\
& & 203\,608 & +0.02 & +0.00 & +0.01 & $-$0.01 & $-$0.01\\
\hline & & 160\,691 & +0.01 & +0.01 & +0.01 & +0.01 & +0.00\\
& & 22\,484 & +0.01 & +0.01 & +0.01 & +0.01 & +0.00\\
\raisebox{1.5ex}[0pt]{$\log g$} & \raisebox{1.5ex}[0pt]{+0.02~dex}
& 63\,077 &
+0.01 & +0.01 & +0.01 & +0.01 & +0.00\\
& & 203\,608 & +0.01 & +0.01 & +0.01 & +0.01 & +0.00\\
\hline & & 160\,691 & +0.00 & +0.01 & +0.00 & +0.01 & +0.00\\
& \raisebox{1.5ex}[0pt]{+0.05 km s$^{-1}$} &
22\,484 & +0.00 & +0.01 & +0.00 & +0.01 & +0.00\\
\cline{2-2} \raisebox{1.5ex}[0pt]{$\xi$} & & 63\,077 & +0.00 &
+0.03 & +0.00
& +0.03 & +0.00\\
& \raisebox{1.5ex}[0pt]{+0.23 km s$^{-1}$} &
203\,608 & +0.00 & +0.03 & +0.00 & +0.03 & +0.00\\
\hline
 & & 160\,691 & +0.03 & +0.02 & +0.04 & +0.03 & +0.01\\
& & 22\,484 & +0.02 & +0.02 & +0.03 & +0.03 & +0.01\\
\raisebox{1.5ex}[0pt]{[Fe/H]} & \raisebox{1.5ex}[0pt]{+0.10~dex} &
63\,077 & +0.02 & +0.02 &
 +0.02 & +0.02 & +0.00\\
& & 203\,608 & +0.01 & +0.01 & +0.02 & +0.02 & +0.01\\
\hline  & & 160\,691 & +0.02 & $-$0.04 & +0.09 & +0.03 & +0.07\\
& \raisebox{1.5ex}[0pt]{$+2\%$} &
22\,484 & +0.03 & $-$0.03 & +0.10 & +0.04 & +0.07\\
\cline{2-2} \raisebox{1.5ex}[0pt]{Continuum} & & 63\,077 & +0.02 &
$-$0.05 & +0.11
& +0.04 & +0.09\\
& \raisebox{1.5ex}[0pt]{$+1\%$} &
203\,608 & +0.03 & $-$0.03 & +0.14 & +0.08 & +0.11\\
\hline \hline
 & & 160\,691 & 0.04 & 0.05 & 0.10 & 0.05 & 0.07\\
\raisebox{-0.3ex}[0pt]{Total} & & 22\,484 & 0.04 & 0.04 & 0.11 & 0.05 & 0.07\\
\raisebox{0.3ex}[0pt]{uncertainty} & & 63\,077 & 0.03 & 0.06 & 0.11 & 0.06 & 0.09\\
 & & 203\,608 & 0.04 & 0.04 & 0.14 & 0.09 & 0.11\\
\hline
\end{tabular}
\end{table*}

\begin{table*}
\caption[]{[Eu/H], [Th/H], and [Th/Eu] abundance ratios for all
sample stars, with their respective uncertainties relative to H
and Fe.} \label{tab:th_eu_abundances}
\begin{tabular} { l c c c c c c c c }
\hline\hline HD & [Eu/H] & uncert.$_{\mathrm{[Eu/H]}}$ &
uncert.$_{\mathrm{[Eu/Fe]}}$ & [Th/H] &
uncert.$_{\mathrm{[Th/H]}}$
& uncert.$_{\mathrm{[Th/Fe]}}$ & [Th/Eu] & uncert.$_{\mathrm{[Th/Eu]}}$\\
\hline
2151 & $-$0.04 & 0.04 & 0.05 & +0.00 & 0.11 & 0.06 & +0.04 & 0.08\\
9562 & +0.05 & 0.04 & 0.05 & +0.04 & 0.11 & 0.06 & $-$0.01 & 0.08\\
16\,417 & +0.12 & 0.04 & 0.05 & +0.06 & 0.11 & 0.06 & $-$0.06 & 0.08\\
20\,766 & $-$0.15 & 0.04 & 0.05 & $-$0.25 & 0.11 & 0.06 & $-$0.10 & 0.08\\
20\,807 & $-$0.12 & 0.04 & 0.05 & $-$0.17 & 0.11 & 0.06 & $-$0.05 & 0.08\\
22\,484 & $-$0.01 & 0.04 & 0.04 & +0.04 & 0.11 & 0.05 & +0.05 & 0.07\\
22\,879 & $-$0.42 & 0.04 & 0.05 & $-$0.58 & 0.12 & 0.07 & $-$0.16 & 0.09\\
30\,562 & +0.23 & 0.04 & 0.05 & +0.19 & 0.11 & 0.06 & $-$0.04 & 0.08\\
43\,947 & $-$0.18 & 0.04 & 0.05 & $-$0.11 & 0.12 & 0.06 & +0.07 & 0.08\\
52\,298 & $-$0.13 & 0.04 & 0.04 & $-$0.09 & 0.12 & 0.06 & +0.04 & 0.09\\
59\,984 & $-$0.48 & 0.04 & 0.05 & $-$0.42 & 0.12 & 0.07 & +0.06 & 0.09\\
63\,077 & $-$0.48 & 0.03 & 0.06 & $-$0.48 & 0.11 & 0.06 & +0.00 & 0.09\\
76\,932 & $-$0.48 & 0.04 & 0.05 & $-$0.54 & 0.12 & 0.07 & $-$0.06 & 0.09\\
102\,365 & $-$0.11 & 0.04 & 0.05 & $-$0.23 & 0.11 & 0.06 & $-$0.12 & 0.08\\
128\,620 & +0.16 & 0.04 & 0.05 & +0.09 & 0.11 & 0.06 & $-$0.07 & 0.08\\
131\,117 & +0.07 & 0.04 & 0.05 & +0.10 & 0.11 & 0.06 & +0.03 & 0.08\\
160\,691 & +0.16 & 0.04 & 0.05 & +0.09 & 0.10 & 0.05 & $-$0.06 & 0.07\\
196\,378 & $-$0.22 & 0.04 & 0.04 & $-$0.31 & 0.12 & 0.07 & $-$0.10 & 0.09\\
199\,288 & $-$0.30 & 0.03 & 0.05 & $-$0.32 & 0.11 & 0.06 & $-$0.02 & 0.09\\
203\,608 & $-$0.50 & 0.04 & 0.04 & $-$0.46 & 0.14 & 0.09 & +0.04 & 0.11\\
\hline
\end{tabular}
\end{table*}

\subsubsection{Comparison with literature results}

There are only two works available in the literature in which Th
abundances have been determined for Galactic disk stars:
\citet{dasilvaetal90} and MKB92. In the classic work of
\citet{butcher87}, which was the first to propose the use of
stellar Th abundances as chronometers, the ratio between the EWs
of Th and Nd is adopted as an estimate of [Th/Nd], but no
independent Th abundances are derived. We did not compare our
results to those from \citeauthor{dasilvaetal90} because only
preliminary data is presented for a small sample of four stars.

\begin{figure}
\resizebox{\hsize}{!}{\includegraphics*{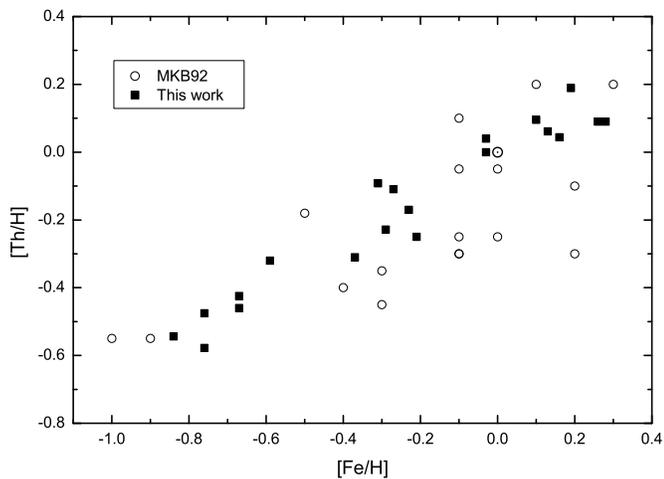}}
\caption{[Th/H] vs. [Fe/H] diagram for our sample stars and those
from MKB92. Note that the behaviour of both data sets is similar,
but that our abundances present considerably lower scatter.}
\label{fig:th_h_fe_h_mkb92}
\end{figure}

MKB92 is a re-analysis of the \citet{butcher87} data, with the
same stellar sample and spectra, but making use of more up-to-date
model atmospheres and more detailed spectral syntheses.
Unfortunately, MKB92 do not present a detailed Th abundance
uncertainty analysis. The authors investigated the influence of
effective temperature, surface gravity and continuum placement
variations, but overlooked metallicity and microturbulence
velocity. No estimate of total uncertainty is presented.
Therefore, we decided not to include error bars in
Fig.~\ref{fig:th_h_fe_h_mkb92}, which depicts [Th/H] vs. [Fe/H]
for our results and those from MKB92. The two data sets exhibit a
similar behaviour, but our results present scatter 61\% lower than
those of MKB92 (from linear fits we get
$\sigma_{\mathrm{our~work}}=0.064$ and
$\sigma_{\mathrm{MKB92}}=0.165$). The lower scatter of our results
is a consequence of many enhancements introduced in our analysis,
like the use of homogeneous and precise atmospheric parameters
obtained by us, while MKB92 used multiple literature sources, and
refinements in the spectral synthesis.

\section{Conclusions}

Homogenous, fully self-consistent atmospheric parameters, and
their respective uncertainties, have been obtained for all sample
stars. The use of two different and internally homogeneous
criteria for effective temperature determination allowed us to
achieve a very low uncertainty (27~K) for this important
parameter. Uncertainties of the other parameters were also found
to be adequate for our needs.

Abundances of the elements that contaminate the \ion{Th}{ii} and
\ion{Eu}{ii} spectral regions -- Ti, V, Cr, Mn, Co, Ni, Ce, Nd and
Sm -- have been determined by detailed spectroscopic analysis,
relative to the Sun, using EWs. For the elements with meaningful
HFSs (V, Mn, and Co), this has been taken into account. A thorough
estimation of the uncertainties has been carried out. The average
[element/H] uncertainty -- $(0.10\pm0.02)$~dex -- was found to be
satisfactorily low.

Eu and Th abundances have been determined for all sample stars
using spectral synthesis of the \ion{Eu}{ii} line at 4129.72~\AA\
and the \ion{Th}{ii} line at 4019.13~\AA. Comparison of our
results with the literature shows that our analysis yielded a
similar behaviour, but with considerably lower scatters (36\%
lower for Eu, and 61\% lower for Th). The [Th/Eu] abundance ratios
thus obtained were used to determine the age of the Galactic disk
in Paper~II.


\begin{acknowledgements}
This paper is based on the PhD thesis of one of the authors
\citep{delpeloso03}. The authors wish to thank the staff of the
Observat\'orio do Pico dos Dias, LNA/MCT, Brazil and of the
European Southern Observatory, La Silla, Chile. The support of
Martin K\"urster (Th\"uringer Landessternwarte Tautenburg,
Germany) during the observations with the ESO's 3.60~m telescope
was greatly appreciated. We thank R. de la Reza, C. Quireza and
S.D. Magalh\~aes for their contributions to this work. EFP
acknowledges financial support from CAPES/PROAP and FAPERJ/FP
(grant E-26/150.567/2003). LS thanks the CNPq, Brazilian Agency,
for the financial support 453529.0.1 and for the grants
301376/86-7 and 304134-2003.1. GFPM acknowledges financial support
from CNPq/Conte\'udos Digitais, CNPq/Institutos do
Mil\^enio/MEGALIT, FINEP/PRONEX (grant 41.96.0908.00), and
FAPESP/Tem\'aticos (grant 00/06769-4). Finally, we are grateful to
the anonymous referee for a thorough revision of the manuscript,
and for comments that helped to greatly enhance the final version
of the work.
\end{acknowledgements}

\bibliographystyle{aa}
\bibliography{referencias}

\end{document}